\documentclass[prb,aps,notitlepage,longbibliography,superscriptaddress]{revtex4-1}
\usepackage{graphicx}
\usepackage{overpic}
\usepackage{subfig}
\usepackage{placeins}
\usepackage{amsmath,amsfonts,amssymb}
\usepackage{mathtools}
\usepackage{units}
\usepackage{tikz}
\usetikzlibrary{arrows.meta}
\usepackage{hyperref}
\usepackage[utf8]{inputenc}
\usepackage[T1]{fontenc}
\usepackage{lmodern}

\newcommand*{\ie}{\mbox{i.\,e.}} 
\newcommand*{\eg}{\mbox{e.\,g.}}

\newcommand*{\td}{\mathrm{d}}
\newcommand*{\pd}{\partial}

\newcommand*{\T}{\mathsf{T}}

\newcommand*{\e}{\mathrm{e}}
\newcommand*{\I}{\mathrm{i}}
\newcommand*{\nn}{\nonumber \\}

\newcommand*{\lft}{\!\left}
\newcommand*{\0}{{\phantom{\dag}}}
\newcommand*{\pI}{{\vphantom{\int}}}

\newcommand*{\pvec}{\vec{p}}
\newcommand*{\sigvec}{\vec{\sigma}}
\newcommand*{\phiL}{\phi_\text{L}}
\newcommand*{\phiR}{\phi_\text{R}}

\newcommand*{\ii}{\mathcal{I}}
\newcommand*{\iiJ}{\ii_\text{J}}
\newcommand*{\Z}{\mathcal{Z}}

\newcommand*{\psup}{\psi_\uparrow}
\newcommand*{\psdn}{\psi_\downarrow}
\newcommand*{\gauge}{\alpha}
\newcommand*{\Dgauge}{\frac{\pd\gauge(x)}{\pd x}}
\newcommand*{\DDgauge}{\frac{\pd^2\gauge(x)}{\pd\phi \, \pd x}}

\newcommand*{\up}{\;\!\uparrow\;\!}
\newcommand*{\subbf}[1]{\textbf{\protect\subref{#1}}}
\newcommand*{\Rcite}[1]{\mbox{Ref.~\onlinecite{#1}}}
\newcommand*{\Rscite}[1]{\mbox{Refs.~\onlinecite{#1}}}
\newcommand*{\myfrac}[2]{%
	{\scriptstyle \nicefrac{\scriptstyle #1}{\scriptstyle #2}}%
}
\makeatletter
\newcommand*{\bigger}{\bBigg@{4}}
\makeatother

\begin{document}

\title{Current--phase relation in a topological Josephson junction: Andreev bands vs.\ scattering states}
\author{Stefan \surname{Backens}}
\affiliation{Institute for Theoretical Condensed Matter physics, Karlsruhe Institute of Technology, D-76131 Karlsruhe, Germany}
\affiliation{Institute for Quantum Materials and Technologies, Karlsruhe Institute of Technology, Hermann-von-Helmholtz-Platz 1, D-76344 Eggenstein-Leopoldshafen, Germany}
\author{Alexander \surname{Shnirman}}
\affiliation{Institute for Theoretical Condensed Matter physics, Karlsruhe Institute of Technology, D-76131 Karlsruhe, Germany}
\affiliation{Institute for Quantum Materials and Technologies, Karlsruhe Institute of Technology, Hermann-von-Helmholtz-Platz 1, D-76344 Eggenstein-Leopoldshafen, Germany}

\date{\today}

\begin{abstract}
We consider a long topological Josephson junction formed on a conducting 2D surface of a 3D topological insulator (TI). 
The superconducting correlations are proximity-induced by s-wave superconductors covering the surface. 
The 1D spacing between the coverings is either unfilled or filled by a magnetic insulator. Generally, the Josephson current mediated by the TI surface is determined by scattering modes as well as by the states localized around the junction ({\it Andreev bound states} or {\it Andreev bands}). We find out that it is crucial to take into account both contributions to 
determine the current--phase relation of the topological Josephson junction. We analyze the dependence of the 
Josephson current on the thickness of the junction as well as the deviations from the sinusoidal shape of the current--phase relation.
\end{abstract}

\maketitle

\section{Introduction}

A conducting surface of a strong 3D topological insulator can be gapped either by a superconducting or a ferromagnetic 
proximity covering. Interfaces between such coverings were predicted to permit the realization of (both zero-dimensional and one-dimensional) Majorana modes by Fu and Kane~\cite{FuKanePRL2008,FuKanePRL2009}. In particular, a narrow 
uncovered or a narrow ferromagnetic stripe between two superconducting coverings gives rise to a long topological Josephson junction. A similar situation also arises in proximitized graphene, as shown in earlier calculations of the Josephson effect~\cite{TitovBeenakkerPRB2006} and of the 1D modes~\cite{TitovOssipovBeenakkerPRB2007}.
Several works have since analyzed the properties of topological Josephson junctions on a TI surface.
For the case of a ferromagnetic insulator separating the superconductors [superconductor--magnetic--superconductor 
(SMS) geometry], the Josephson current resulting from Andreev \emph{bound states}, in the form of chiral Majorana modes, has been calculated for s-wave superconductors \cite{TanakaPRL2009} and for different superconducting pairings \cite{LinderPRB2010}; both a ferromagnetic and an ungapped middle region [superconductor--normal metal--superconductor (SNS) case] were considered in \Rcite{SnelderPRB2013}. 
Recently, a Josephson Hall current, perpendicular to the regular Josephson current, was derived from \emph{scattering-state} wave functions in an SMS system~\cite{Maistrenko2020}. Closely related physics has been studied both theoretically \cite{Pientka2017} and experimentally 
\cite{Fornieri2019,Ren2019} in 2D semiconducting films with strong spin--orbit coupling covered by 
superconductors.

Theoretical analysis so far has focused on the case of a large chemical potential~$\mu$ in comparison to the superconducting gap on the TI surface, at least in the areas covered by superconductors~\cite{TitovBeenakkerPRB2006,TanakaPRL2009,LinderPRB2010,SnelderPRB2013,Maistrenko2020}. 
This case currently seems to be of most experimental relevance. 
In this article, we analyze  a topological Josephson junction in the---presently at least theoretically---interesting limit of an overall vanishing chemical potential, \ie, at the Dirac point. We choose two relatively simple examples: 1) SNS contact; 2) SMS contact with the ferromagnetic stripe polarized in $z$~direction and with the ferromagnetic energy gap being equal to the superconducting one. Our aim is not to treat the most general magnetic configurations 
(see, \eg, \Rscite{Maistrenko2020,TanakaPRL2009}) but rather to consider two simple cases where we  
can fully take into account the contributions of the Andreev bound states and the scattering states to the Josephson current. We also determine the asymptotic behavior of the Josephson 
current in the limits of both large and small thickness of the junction. In contrast to the case of a high chemical potential 
in the parts covered by the superconductor~\cite{TitovBeenakkerPRB2006}, the behavior at the Dirac point does \emph{not} conform to the Ambegaokar--Baratoff relation.

\section{The system}

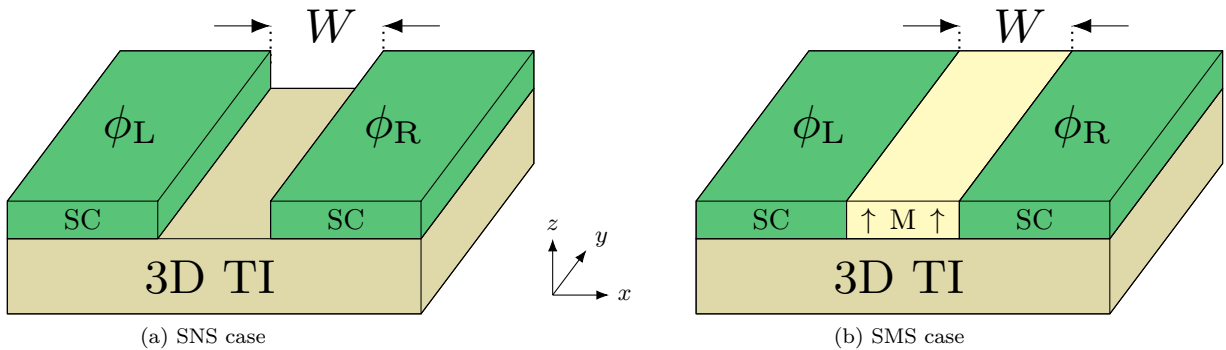
\begin{figure}
\centering
\definecolor{gruen}{RGB}{89,196,119}
\newcommand*{\zdimx}{1.5}
\newcommand*{\zdimy}{2}
\newcommand*{\schicht}{0.5}
\pgfmathsetmacro{\hoehe}{2*\schicht}
\newcommand*{\seite}{2}
\newcommand*{\halb}{0.75}
\pgfmathsetmacro{\breite}{2*\halb}
\pgfmathsetmacro{\haelfte}{\seite + \halb}
\pgfmathsetmacro{\gesamt}{2*\haelfte}
\pgfmathsetmacro{\seitM}{\halb + 0.5*\seite}
\newcommand*{\quadlin}[3]{
	\draw (#3) rectangle ++(#1, #2) -- +(\zdimx, \zdimy) ++(0, -#2)
		-- ++(\zdimx, \zdimy) -- ++(0, #2) -- ++(-#1, 0) -- ++(-\zdimx, -\zdimy);
}
\newcommand*{\quadfill}[4]{
	\filldraw[#4] (#3) -- ++(0, #2) -- ++(\zdimx, \zdimy) -- ++(#1, 0)
		-- ++(0, -#2) -- ++(-\zdimx, -\zdimy) -- ++(-#1, 0);
}
\newcommand*{\TImitB}{
	\quadfill{\gesamt}{\hoehe}{-\haelfte, -\hoehe}{fill=white!70!olive}
	\quadlin{\gesamt}{\hoehe}{-\haelfte, -\hoehe}
	\path (0, -\schicht) node {\scalebox{2}{3D TI\,}};

	\path (\zdimx, \zdimy) ++(0, \schicht) node[above] (breite) {\scalebox{2}{$\,W$}};
	\draw[arrows={->[scale=1.5]}] (breite) +(-\breite, 0) -- +(-\halb, 0);
	\draw[arrows={->[scale=1.5]}] (breite) +(\breite, 0) -- +(\halb, 0);
	\draw[dotted,thick] (breite) +(-\halb, -\schicht) -- +(-\halb, 0)
		+(\halb, -\schicht) -- +(\halb, 0);
}
\newcommand*{\einSC}[3]{
	\quadfill{\seite}{\schicht}{#1, 0}{fill=gruen}
	\quadlin{\seite}{\schicht}{#1, 0}
	\path (#2, \schicht) node[below] {\scalebox{1.2}{SC}} ++(\halb, \hoehe) node {\scalebox{2}{#3}};
}
\hspace{-2.6cm}
\subfloat[SNS case]{
	\hspace{1.6cm}
	\begin{tikzpicture}[>=Latex]
	\TImitB
	\einSC{-\haelfte}{-\seitM}{$\phiL~$}
	\einSC{\halb}{\seitM}{$\phiR~$}

	\draw[->] (4.5, -\halb) -- +(\halb, 0) node[right] {$x$};
	\draw[->] (4.5, -\halb) -- +(0.45, 0.6) coordinate (yy);
	\path (yy) +(0.2, 0.15) node {$y$};
	\draw[->] (4.5, -\halb) -- +(0, \halb) node[above] {$z$};
	\end{tikzpicture}
	\hspace{-1.6cm}
\label{J-geoSNS}}
\hspace{1cm}
\subfloat[SMS case]{
	\hspace{0.8cm}
	\begin{tikzpicture}[>=Latex]
	\TImitB
	\einSC{-\haelfte}{-\seitM}{$\phiL~$}

	\quadfill{\breite}{\schicht}{-\halb, 0}{fill=white!70!yellow}
	\quadlin{\breite}{\schicht}{-\halb, 0}
	\path (0, \schicht) node {\raisebox{-8mm}{\scalebox{1.2}{$\up \phantom{M} \up$}}};
	\path (0, \schicht) node[below] {\scalebox{1.2}{M}};

	\einSC{\halb}{\seitM}{$\phiR~$}
	\end{tikzpicture}
	\hspace{-0.8cm}
\label{J-geoSMS}}
\caption{Two s-wave superconductors (SC) with phases $\phiL$, $\phiR$ form a long Josephson junction of width~$W$ on the surface of a three-dimensional topological insulator (3D TI). 
In the middle, \subbf{J-geoSNS} the TI surface either remains ungapped, or \subbf{J-geoSMS} it is gapped by a ferromagnetic insulator (M) with magnetization perpendicular to the surface. 
}\label{J-geo}
\end{figure}

The two systems studied are shown in Fig.~\ref{J-geo}. 
The Hamiltonian governing the surface of the TI is given by
\begin{subequations}\label{NambuH}
\begin{align}
H &= \frac{1}{2} \int\! \td x \, \td y \, \Psi^\dag \, h \, \Psi~, \\
h &= v \, \pvec \cdot \sigvec \, \tau_z + M(x) \, \sigma_z
	+ \Delta(x) \, \tau_{-} + \Delta^*(x) \, \tau_{+}~,
\end{align}
\end{subequations}
with the fermion fields $\Psi = [\psup, \psdn, \psdn^\dag, -\psup^\dag]^\T$, the Fermi velocity $v > 0$ and the 2D momentum operator $\pvec = -\I \, \hbar \, \vec \nabla$. 
Pauli matrices $\sigma_j$ and $\tau_j$ act in spin and Nambu space, respectively. The chemical potential is 
chosen to vanish everywhere.
We consider the superconducting order parameter to be of the same magnitude $\Delta_0 > 0$ on both sides,
\begin{align}
\Delta(x) &= \Delta_0 \, \left[ \Theta(-x - W/2) \, \e^{\I \, \phiL} + \Theta(x - W/2) \, \e^{\I \, \phiR} \right]~,
\intertext{where $W > 0$ and $\Theta$ is the Heaviside function, and the phase difference $\phi \equiv \phiL - \phiR$
is arbitrary. In the SMS case, the strip 
between the superconductors is filled by a ferromagnetic insulator inducing a magnetic gap}
M(x) &= M_0 \, \Theta(W/2 - |x|)\ .
\end{align}
For simplicity we choose $M_0 = \Delta_0$ in the SMS case, so that the superconducting and the magnetic gaps are equal in amplitude. The SNS case corresponds trivially to $M_0=0$.

\section{Formulation of the problem}

In multiple previous studies of topological Josephson junctions\cite{FuKanePRL2008,TanakaPRL2009,LinderPRB2010,SnelderPRB2013,Pientka2017}, the principal role was attributed to the 1D ``bound states''. These are, actually, 1D continuum bands of Andreev states propagating along the junction with energies lying in the gap of the corresponding 2D continuum. These states are bound only in the direction perpendicular to that of the long junction. It is frequently assumed that 1D states are mostly responsible for the Josephson current (see however \Rcite{Maistrenko2020}). 

\begin{figure}[t]
\begin{overpic}[height=5cm]{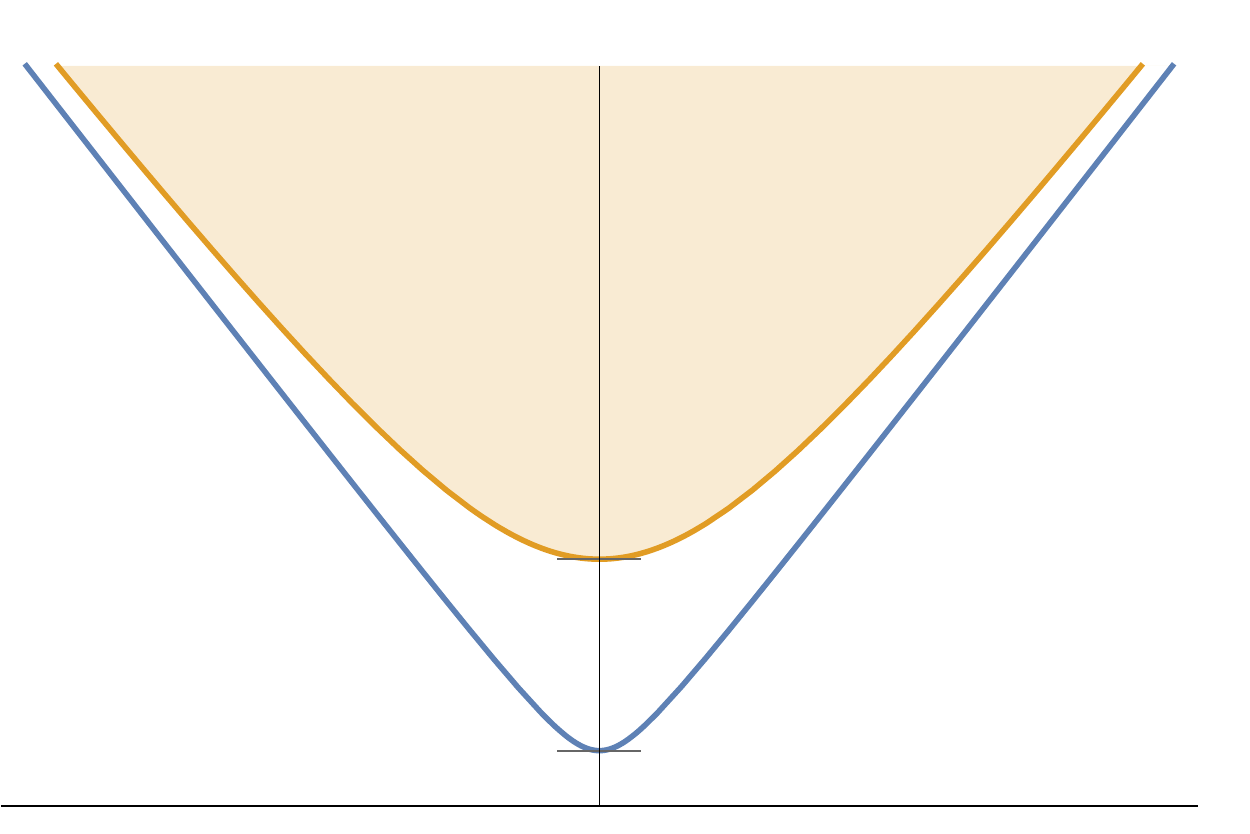}
	\put(97.5,1){$q$}
	\put(47.25,63){$\epsilon$}
	\put(38.5,20.25){$\Delta_0$}
	\put(39.5,5.5){$m$}
\end{overpic}
\caption{Continuum of 2D states (shaded area) and \emph{Andreev band} of 1D states (blue curve). 
The Andreev states are localized in the direction perpendicular to the long junction, but propogate freely, with momentum~$q$, along the junction. 
While the 1D states lie in the gap of the 2D continuum for any \emph{fixed} value of $q$, they are not restricted to~energies below the superconducting gap~$\Delta_0$. 
}\label{J-bands}
\end{figure}

Let us look closer at such states in our case.
Since the Hamiltonian \eqref{NambuH} is homogeneous in $y$~direction, all states are characterized 
by the $y$~component of the momentum, $q \equiv p_y$. The dispersion relation of the 2D continuum 
(scattering states) is given by $\epsilon(k,q) = \sqrt{\Delta_0^2 + v^2 \, (k^2 + q^2)}$, where $k\equiv p_x$ 
is the $x$~component of the momentum of the incoming (and the outgoing) waves. In contrast, the 1D states are characterized 
by the dispersion relation $\epsilon_m(q) =\sqrt{m^2 + v^2 \, q^2}$ with the phase dependent mass $m(\phi) < \Delta_0$. At any given value of $q$, the 
energy of the 1D states is below the lower band edge of the 2D states, 
$\epsilon_m(q) < \epsilon(k,q)$. In this sense, the 1D states lie in the gap, and they are bound states which form \emph{Andreev bands}. 
However, $\epsilon_m(q)$ is not limited 
by $\Delta_0$ (see Fig.~\ref{J-bands}). 
The simplest case arises for the SMS junction. For $W\rightarrow \infty$, there are two decoupled chiral Majorana modes $\chi_L$ and $\chi_R$ localized near the two SM interfaces.\cite{FuKanePRL2008,FuKanePRL2009} With decreasing $W$, these modes hybridize and the effective Hamiltonian reads
\begin{align}
H_{\rm eff} = \frac{1}{2} \int\! \td y \ \big(\chi_L\ \chi_R\big)
	\left(\begin{array}{cc} -\I \, \hbar \, v \, \partial_y & \I \, m \\
-\I \, m & \I \, \hbar \, v \, \partial_y \end{array}\right)\left(\begin{array}{c}
	\chi_L \\ \chi_R\end{array}\right)\ ,
\end{align}
where 
\begin{align}\label{eq:mass0W}
m(\phi) = m_0 \cos\lft(\frac{\phi}{2}\right)\ .
\end{align}
The single-particle spectrum of $H_{\rm eff}$ is given by $\epsilon_m(q) =\sqrt{m^2 + v^2 \, q^2}$.
For $W \ll \xi$, one obtains~\cite{FuKanePRL2008} $m_0 \rightarrow \Delta_0$.

Considering exclusively the Andreev (bound) states, the standard argument (cf.~App.~\ref{appSuperEigen}) leads to the following result for the Josephson current 
\begin{align}
\iiJ = L_y \int\! \frac{\td q}{2 \pi \, \hbar} \left(\frac{2 \, e}{\hbar} \, \frac{\partial \epsilon_m}{\partial \phi}\right) \left[-\frac{1}{2} \, \tanh\lft(\frac{\epsilon_m}{2 \, k_B \, T}\right)\right]\ ,
\end{align}
which gives a logarithmically divergent result:
\begin{align}\label{eq:IJlog}
\iiJ \sim L_y \, \frac{2 \, e}{\hbar} \,\frac{m_0^2}{\hbar \, v} \, \sin(\phi)\,\log\lft(\frac{q_{\rm max}}{\dots}\right)\ .
\end{align}
Motivated by the pre-factor of \eqref{eq:IJlog} and having in mind the limit $m_0 \sim \Delta_0$ we introduce
the characteristic current scale  
\begin{align} \label{defs}
\ii_0 = \frac{e \, L_y \, \Delta_0^2}{\hbar^2 \, v} = \frac{e \, \Delta_0}{\hbar} \,\frac{L_y}{\xi}\ ,
\end{align}
where $\xi \equiv \frac{\hbar \, v}{\Delta_0}$ is the coherence length.

The logarithmic divergence in \eqref{eq:IJlog} is clearly unphysical. It has to do with us neglecting the contribution 
of the scattering states. In what follows, we provide the full calculation, which remedies this deficiency. We will observe 
that the contribution of the scattering states compensates the divergency, so that a cutoff-independent result emerges. 
In the SNS case in the limit of a wide junction ($W > \xi$), the interplay is even more interesting. In this case, there are multiple Andreev bands, some of which merge into the continuum at certain values of the phase bias $\phi$. 
This leads to quite a nontrivial contribution of the Andreev states to the Josephson current, showing, \eg, a $\pi$-junction 
behavior, kinks etc. Yet, all these features are compensated by the contribution of the scattering states.

\section{SNS case}\label{sec:SNS}

\begin{figure}[t]
\hspace{0.9cm}
\subfloat[$W = 0$]{\hspace{-0.5cm}\begin{overpic}[height=5cm]{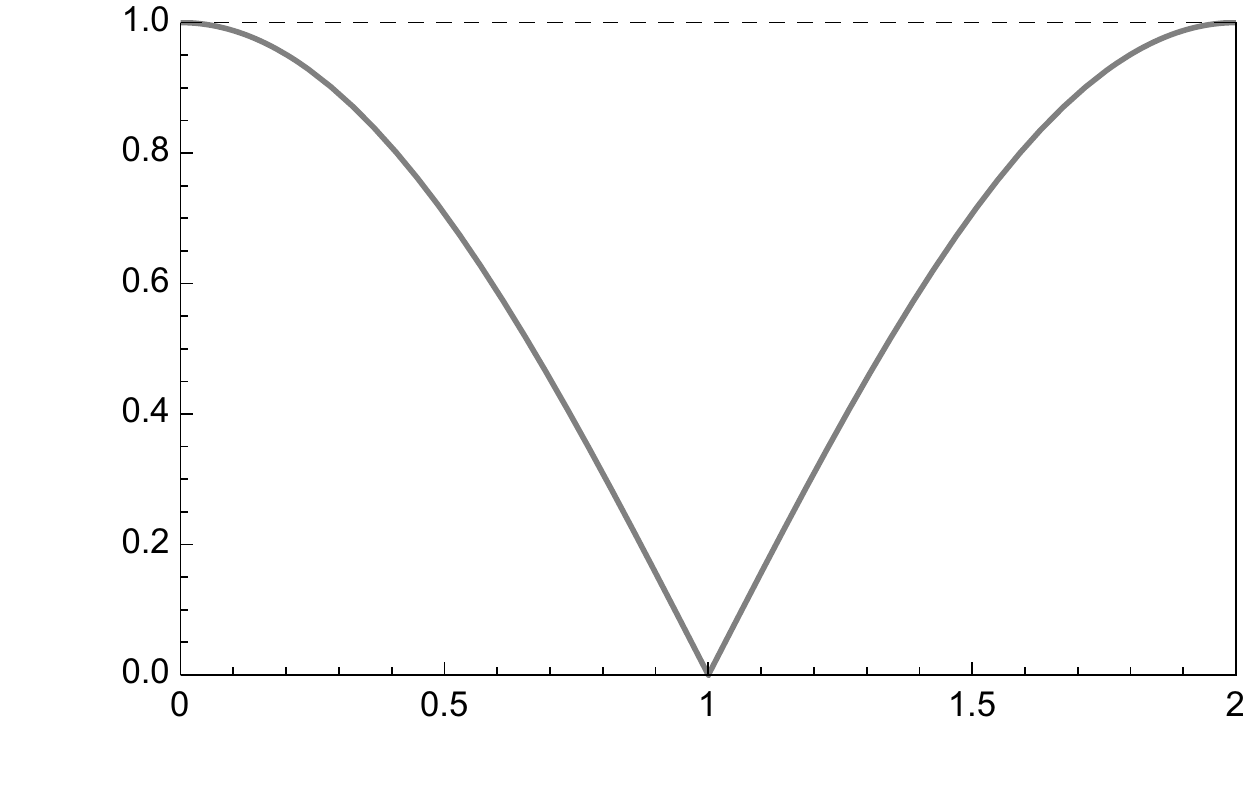}
	\put(52.75,2){\scalebox{1.3}{$\myfrac{\phi\,}{\;\!\pi}$}}
	\put(-5.5,35.5){\scalebox{1.3}{$\frac{|m(\phi)|}{\Delta_0\pI}$}}
\end{overpic}\hspace{0.5cm}\label{J-mSNS0}}
\hspace{0.7cm}
\subfloat[$W = 2.5 \, \xi$]{\hspace{-0.5cm}\begin{overpic}[height=5cm]{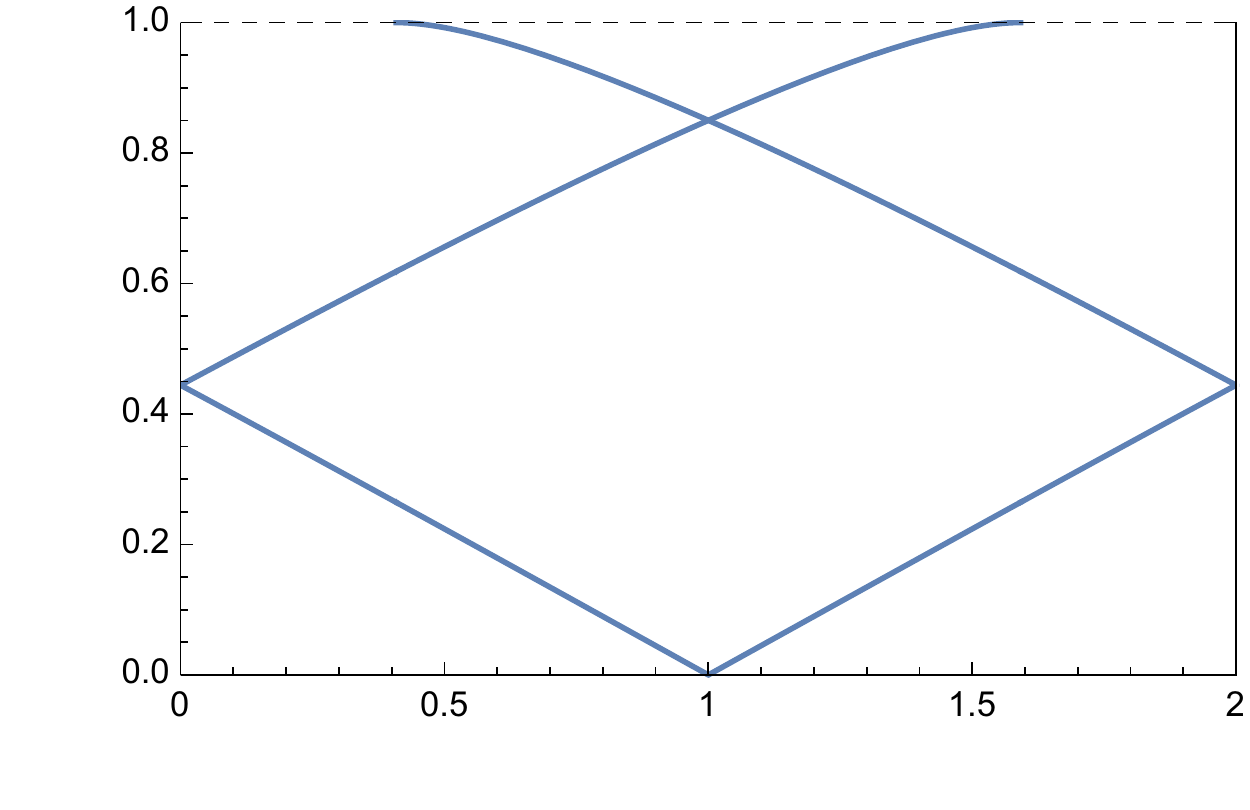}
	\put(52.75,2){\scalebox{1.3}{$\myfrac{\phi\,}{\;\!\pi}$}}
	\put(-5.5,35.5){\scalebox{1.3}{$\frac{|m(\phi)|}{\Delta_0\pI}$}}
\end{overpic}\hspace{0.5cm}\label{J-mSNS25}}
\hspace*{-0.7cm}
\caption{In the SNS case, the number of Andreev bands depends both on the junction width~$W$ and on the phase difference~$\phi$. \subbf{J-mSNS0} In the limit of zero width, the mass is given by $|m(\phi)| = \Delta_0 \, |\cos(\phi/2)|$, cf.~Eq.~\eqref{eq:mass0W}. \\
\subbf{J-mSNS25} For wider junctions, Andreev bands merge with the continuum, $|m(\phi)| \rightarrow \Delta_0$, at certain combinations of $W$ and~$\phi$. 
The effective mass~$m$ is determined by the condition~\eqref{qSNSb} and does not depend on the momentum~$q$ in the~translation-invariant direction. 
As the sign of $m$ has no impact, only the absolute value is shown here. 
}\label{J-mSNS}
\end{figure}

The derivation of the Josephson current is discussed in App.~\ref{appSuperEigen}. The general approach for finding eigenmodes of the Hamiltonian~\eqref{NambuH} is explicated in App.~\ref{appSpectrum}. First, we 
treat the SNS case, $M_0 \equiv 0$. 

{\it Andreev bands.} We start with the contribution of the Andreev states.
The spectrum of the Andreev bands (cf.~App.~\ref{SNSb}) is determined by the following quantization condition:
\begin{align}
\exp\lft[ \I \,\frac{2 \, m \, W}{\hbar \, v} + \I \, \phi \right]
	&= \left( \frac{m + \I \, \sqrt{\Delta_0^2 - m^2}}{\Delta_0} \right)^2~, \label{qSNSb}
\end{align}
which depends on the width of the junction~$W$ and the phase difference~$\phi$ between the superconducting regions. 
In general, there are multiple Andreev bands, which merge with the continuum at certain combinations of $W$ and $\phi$. 
The sign of the mass~$m$ is irrelevant for these calculations: both positive and negative $m$ are allowed here. 
Fig.~\ref{J-mSNS} illustrates the behaviour of the absolute value~$|m(\phi)|$. 

To calculate the Josephson current, we need to sum over solutions~$m_j$ of the condition~\eqref{qSNSb}. 
Every such solution gives rise to a 1D Andreev band with dispersion $\epsilon_j = \sqrt{m_j^2 + v^2 \, q^2}$.
From the quantization condition, we can determine the phase derivative
\begin{align}
\frac{\pd m_j}{\pd\phi} &= -\frac{1}{2}
	\, \frac{\sqrt{\Delta_0^2 - m_j^2}}{1 + \frac{W}{\hbar v} \, \sqrt{\Delta_0^2 - m_j^2}}~, \quad
\end{align}
which appears in the Andreev-state contribution to the Josephson current~\eqref{super.iiJ}:
\begin{align}
\frac{\ii^\text{b}_\zeta}{L_y} &= -\frac{e}{\hbar^2} \sum_{m_j} \int_0^{\zeta/v} \frac{\td q}{\pi}
	\, \frac{\pd\epsilon_j}{\pd\phi} \tanh\lft( \frac{\epsilon_j}{2 \, k_\text{B} \, T} \right) \nn
&= -\frac{e}{\hbar^2} \sum_{m_j} \int_0^{\zeta/v} \frac{\td q}{\pi}
	\, \frac{m_j}{\sqrt{m_j^2 + v^2 \, q^2}} \, \frac{\pd m_j}{\pd\phi}
	\, \tanh\lft( \frac{\sqrt{m_j^2 + v^2 \, q^2}}{2 \, k_\text{B} \, T} \right)~, \label{bSNSii}
\intertext{where we have introduced the $q$ cutoff in units of energy $\zeta \equiv v \, q_\text{max} \gg \Delta_0$. 
For zero-temperature, we obtain the expression}
\frac{\ii^\text{b}_\zeta}{L_y} \:\:&\overset{\mathclap{T=0}}{\approx}\:\: \frac{e}{2 \pi \, \hbar^2 \, v} \sum_{m_j}
	\ln\lft( \frac{2 \, \zeta}{|m_j|} \right) \, \frac{m_j \, \sqrt{\Delta_0^2 - m_j^2}}
	{1 + \frac{W}{\hbar \, v} \, \sqrt{\Delta_0^2 - m_j^2}}~.
\end{align}

{\it Scattering states.} For the scattering states (cf.~App.~\ref{SNSsc}), we integrate the contribution $I_\text{sc}(k, q)$ per state over momenta~$k$, $q$ to obtain the thermal expectation value of the current~\eqref{super.iiJ}:
\begin{align}
\ii^\text{sc}_\zeta &= -\frac{1}{2 \, \hbar^2} \int_0^\infty \frac{L_x}{2 \pi} \, \td k \, \int_0^{\zeta/v}
	\frac{2 L_y}{2 \pi} \: \td q \; I_\text{sc}(k, q) \, \tanh\lft( 
	\frac{\sqrt{\Delta_0^2 + v^2 \, (k^2 + q^2)}}{2 \, k_\text{B} \, T} \right)~. \label{scSNSii}
\intertext{We use the same cutoff as for the Andreev states, $v \, q_\text{max} = \zeta \gg \Delta_0$, to regularize the divergent integral over $q$; the additional factor of 2 replaces the integration over negative values of $q$. 
With the expression for $I_\text{sc}(k, q)$, Eq.~\eqref{IkqSNS}, the $q$ integration can be carried out analytically for zero temperature:}
\frac{\ii^\text{sc}_\zeta}{L_y} \:\:&\overset{\mathclap{T=0}}{=}\:\: -\frac{e}{\pi \, \hbar^2} \int_0^\infty \frac{\td k}{2 \pi}
	\, \sinh^{-1}\lft( \frac{\zeta}{\sqrt{\Delta_0^2 + v^2 \, k^2}} \right) \nn
&\:\:\quad\: \cdot \frac{\Delta_0^2 \; v^2 \, k^2 \, \sqrt{\Delta_0^2 + v^2 \, k^2}
	\, \sin\lft( 2 \, \frac{W}{\hbar} \, \sqrt{k^2 + \frac{\Delta_0^2}{v^2}} \right) \, \sin(\phi)}
	{\left[ v^2 \, k^2 + \Delta_0^2 \, \sin^2\lft( \frac{W}{\hbar} \, \sqrt{k^2 + \frac{\Delta_0^2}{v^2}}
	- \dfrac{\phi}{2} \right) \right] \left[ v^2 \, k^2 + \Delta_0^2 \, \sin^2\lft(
	\frac{W}{\hbar} \, \sqrt{k^2 + \frac{\Delta_0^2}{v^2}} + \dfrac{\phi}{2} \right) \right]}~.
\end{align}

\section{SMS case}\label{sec:SMS}

\begin{figure}[t]
\hspace{0.9cm}
\subfloat[Phase dependence]{\hspace{-0.5cm}\begin{overpic}[height=5cm]{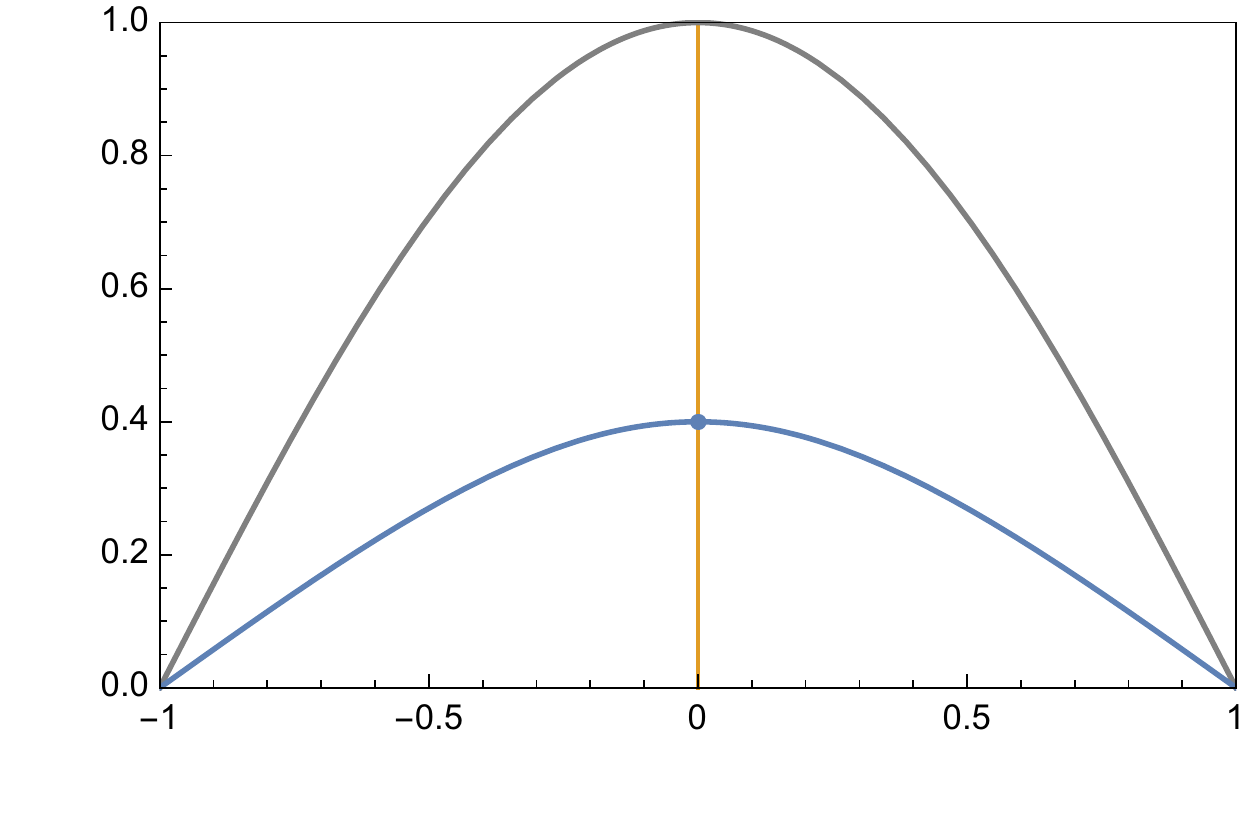}
	\put(52,1.5){\scalebox{1.3}{$\myfrac{\phi\,}{\;\!\pi}$}}
	\put(-5.5,35.5){\scalebox{1.3}{$\frac{m(\phi)}{\Delta_0\pI}$}}
	\put(58.5,52){\scalebox{1.3}{$\scriptstyle W=\,0$}}
	\put(58.5,23.5){\scalebox{1.3}{$\scriptstyle W=\,\xi$}}
\end{overpic}\hspace{0.5cm}\label{J-mSMSphase}}
\hspace{0.8cm}
\subfloat[Width dependence]{\hspace{-0.5cm}\begin{overpic}[height=5cm]{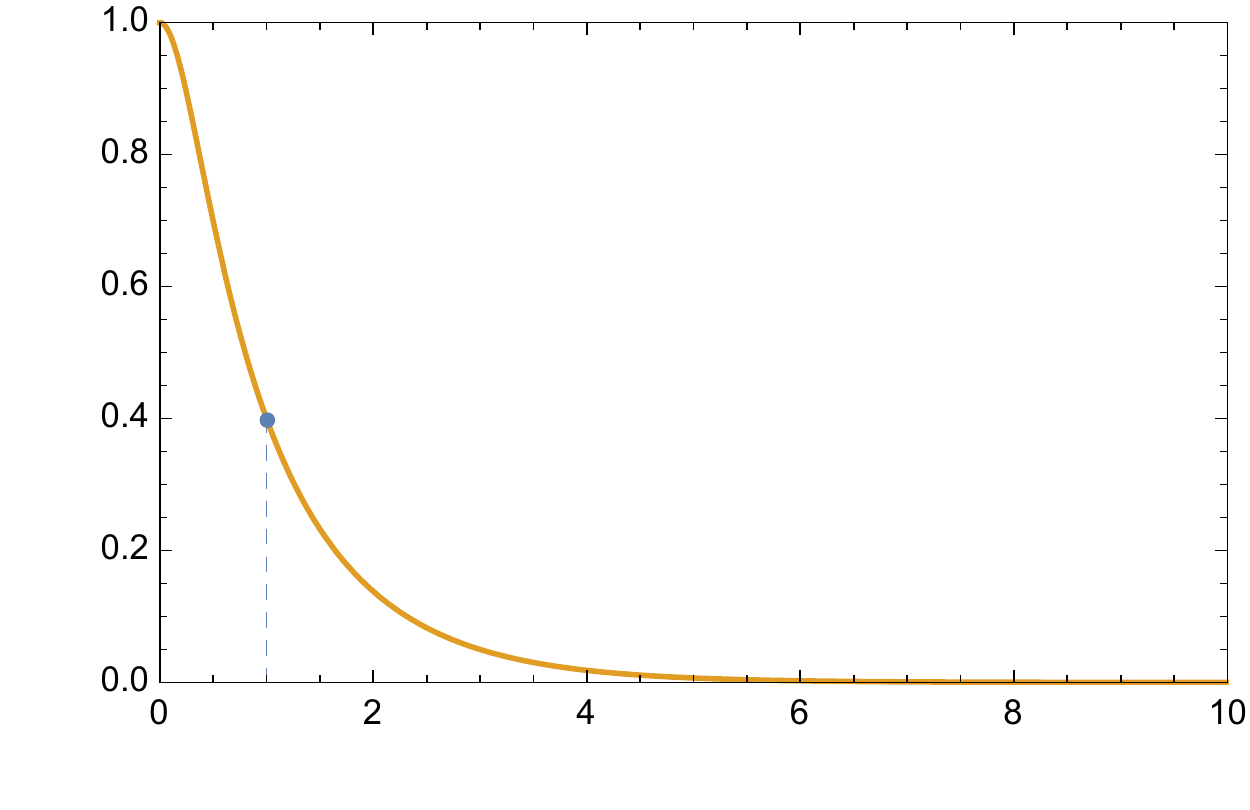}
	\put(51.25,2){\scalebox{1.3}{$\scriptstyle W / \:\!\xi$}}
	\put(-5,35.5){\scalebox{1.3}{$\frac{m(0)}{\Delta_0\pI}$}}
\end{overpic}\hspace{0.5cm}\label{J-mSMSwidth}}
\hspace*{-0.5cm}
\caption{In the SMS case, there is always exactly one Andreev band. 
\subbf{J-mSMSphase} The phase-dependence of the mass~$m$ is unremarkable, although it deviates from the cosine shape for finite junction width. 
\subbf{J-mSMSwidth} The maximum $m(\phi = 0)$ is exponentially suppressed for larger junction widths~$W$. 
The effective mass~$m$ is determined by the condition~\eqref{qSMSb} and does not depend on the momentum~$q$ in the trans\-lation-invariant direction. 
}\label{J-mSMS}
\end{figure}
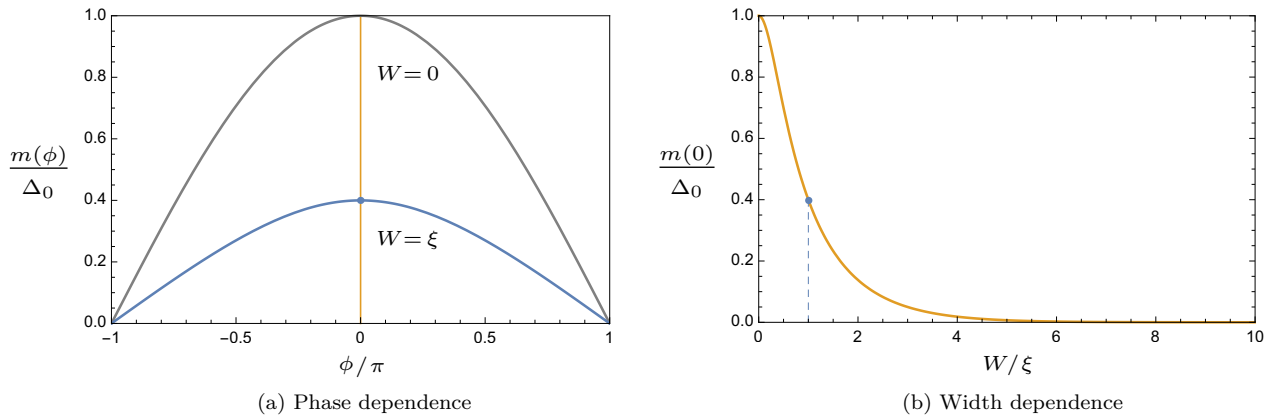

We switch to the SMS case, $M_0 = \Delta_0$, now. The Josephson current is calculated similar to the SNS case of Sec.~\ref{sec:SNS}. 

{\it Andreev bands.}
The quantization condition for the Andreev states (cf.~App.~\ref{SMSb}) is given by
\begin{align}
m &= \Delta_0 \, \cos\lft[ \frac{\phi}{2} \right]
	\, \exp\lft[-\frac{W}{\hbar \, v} \, \sqrt{\Delta_0^2 - m^2} \right]~. \label{qSMSb}
\end{align}
The effective mass~$m$ is uniquely determined by this self-consistency equation: 
The solution~$m(\phi)$ of the quantization condition~\eqref{qSMSb} is a single-valued function for any junction width~$W$ (cf.~Fig.~\ref{J-mSMS}). 
When we insert the derivative
\begin{align}
\frac{\pd m}{\pd\phi} &= \frac{1}{2} \, \frac{m \, \tan\lft[ \frac{\phi}{2} \right]}
	{\frac{W}{\hbar \, v} \, \frac{m^2}{\sqrt{\Delta_0^2 - m^2}} - 1} \phantom{~.}
\end{align}
into the expression~\eqref{bSNSii}, we can write down the contribution of the Andreev band for zero temperature:
\begin{align}
\frac{\ii^\text{b}_\zeta}{L_y} \:\:&\overset{\mathclap{T=0}}{\approx}\:\: \frac{e}{2 \pi \, \hbar^2 \, v}
	\, \ln\lft( \frac{2 \, \zeta}{|m|} \right) \, \frac{m^2 \, \tan\lft[ \frac{\phi}{2} \right]}
	{1 - \frac{W}{\hbar \, v} \, \frac{m^2}{\sqrt{\Delta_0^2 - m^2}}}~. \label{IbSMS}
\end{align}

{\it Scattering states.}
The integration over the scattering-state contributions (cf.~App.~\ref{SMSsc}) is treated essentially like in Eq.~\eqref{scSNSii}. 
For the numerical evaluation in the following section, however, a symmetrical cutoff $\Theta\lft[ \zeta^2 - v^2 \, (k^2 + q^2) \right]$ turns out to be more expedient in this case. 
From the expression~\eqref{IkqSMS}, we obtain the zero-temperature current
\begin{align}
\frac{\ii^\text{sc}_\zeta}{L_y} \:\:&\overset{\mathclap{T=0}}{=}\:\: -\frac{e}{\pi \, \hbar^2}
	\int_0^{\zeta/v} \frac{\td k}{2 \pi} \, \ln\lft( \frac{\sqrt{\Delta_0^2 + \zeta^2}
	+ \sqrt{\zeta^2 - v^2 \, k^2}}{\sqrt{\Delta_0^2 + v^2 \, k^2}} \right) \nn
&\:\:\quad\: \cdot \frac{\Delta_0^2 \; \big( \Delta_0^2 + v^2 \, k^2 \big)
	\, v \, k \, \sin[2 \, W \, k / \hbar] \, \sin[\phi]}
	{\Bigl( v^2 \, k^2 + \Delta_0^2 \, \sin^2\lft[ \frac{\phi}{2} \right] \Bigr)^2 + \, 4 \, \Delta_0^2 \;
	\big( \Delta_0^2 + v^2 \, k^2 \big) \, \cos^2\lft[ \frac{\phi}{2} \right] \, \sin^2[W \, k / \hbar]}~.
\end{align}

\section{Numerical results for $T=0$}\label{zero}

\begin{figure}[t]
\hspace*{0.5cm}
\subfloat[SNS case]{\hspace{-0.5cm}\begin{overpic}[height=5cm]{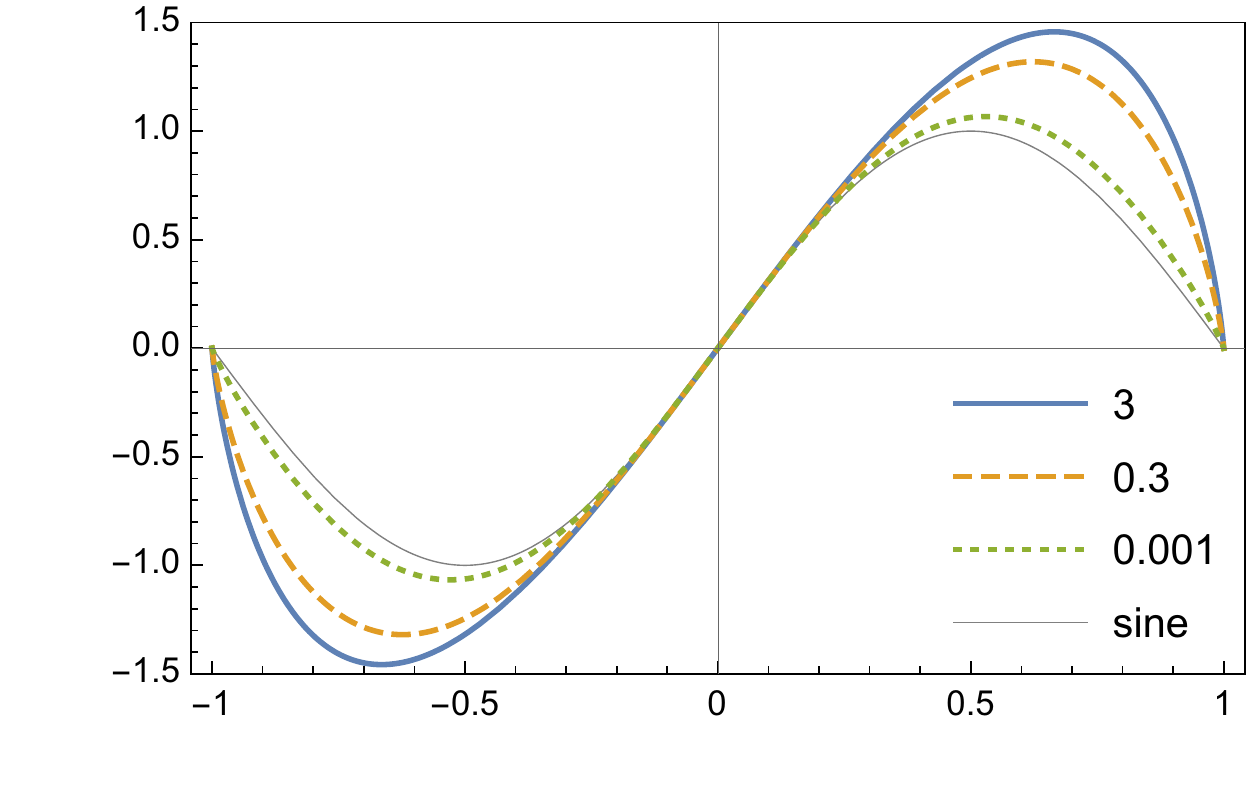}
	\put(53.5,2){\scalebox{1.3}{$\myfrac{\phi\,}{\;\!\pi}$}}
	\put(-3.5,35){\scalebox{1.3}{$\frac{\iiJ(\phi)}{\iiJ'(0)\pI}$}}
	\put(59,24.5){\scalebox{1.3}{$\frac{W}{\xi} \, {=} \: \Bigg\{$}}
\end{overpic}\hspace{0.5cm}\label{J-SNSchi}}
\hspace{0.7cm}
\subfloat[SMS case]{\hspace{-0.5cm}\begin{overpic}[height=5cm]{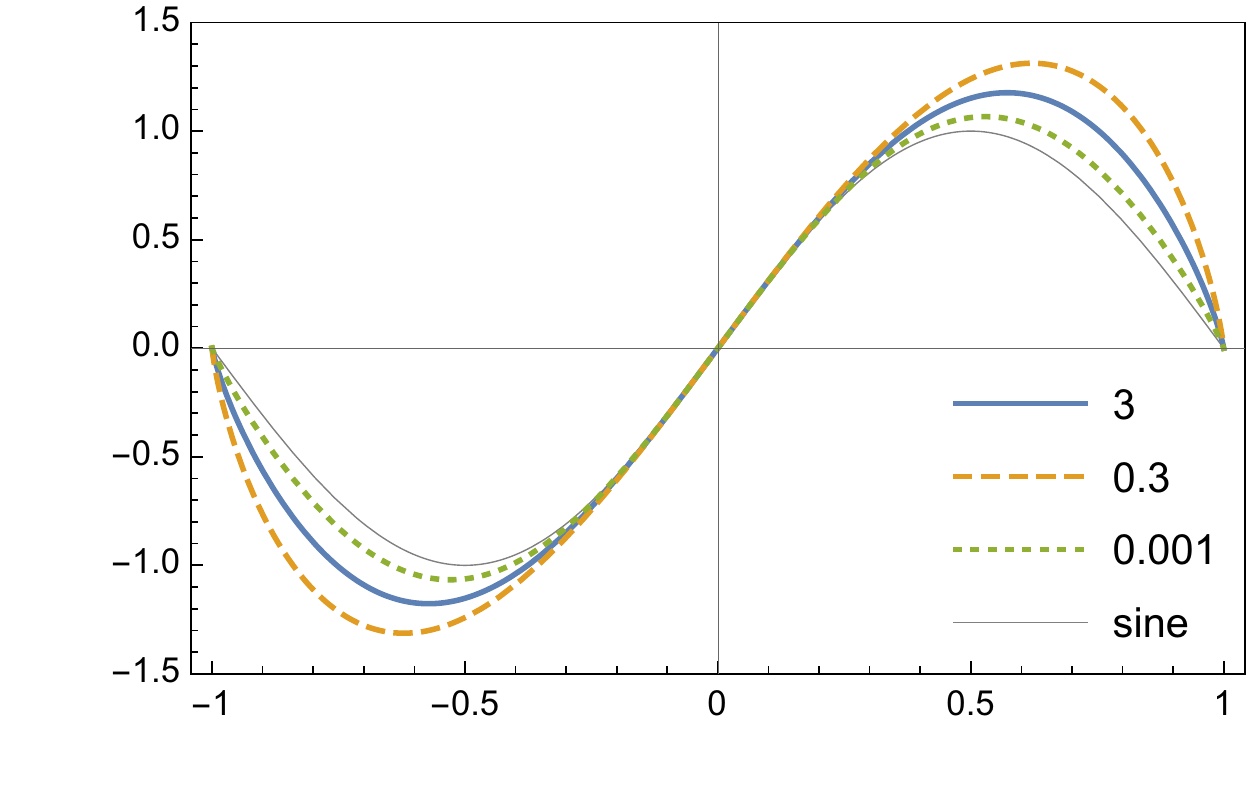}
	\put(53.5,2){\scalebox{1.3}{$\myfrac{\phi\,}{\;\!\pi}$}}
	\put(-3.5,35){\scalebox{1.3}{$\frac{\iiJ(\phi)}{\iiJ'(0)\pI}$}}
	\put(59,24.5){\scalebox{1.3}{$\frac{W}{\xi} \, {=} \: \Bigg\{$}}
\end{overpic}\hspace{0.5cm}\label{J-SMSchi}}
\hspace*{-0.7cm}
\caption{To illustrate the shape of the current--phase relation, curves for different values of the junction width~$W$ (specified in the plot legends) are normalized to a uniform slope at $\phi = 0$. 
Additionally, a \emph{sine} is shown for comparison. 
Note that in the magnetic case~\subbf{J-SMSchi}, the shape changes non-monotonously (cf.~Fig.~\ref{J-ratioPlot}). 
}\label{J-shapePlot}
\end{figure}
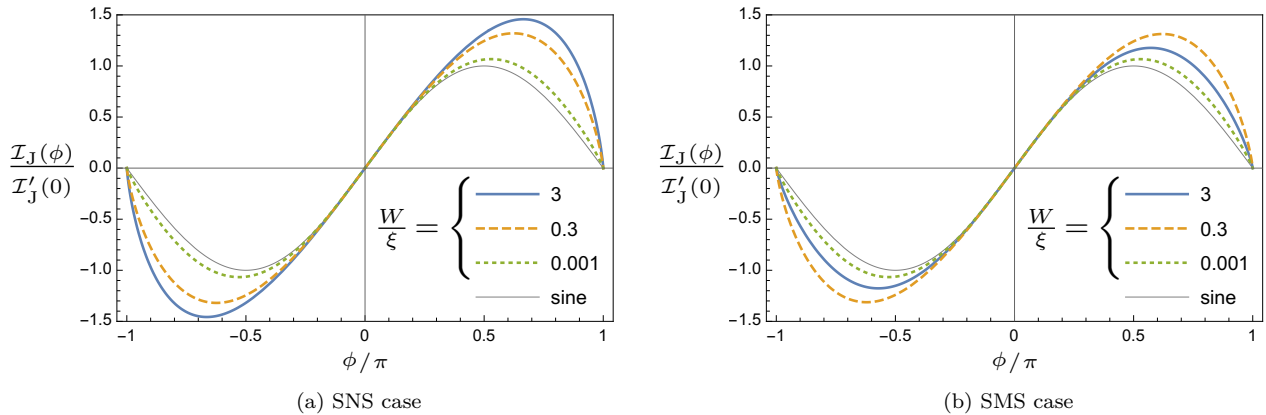

Both the Andreev-state and the scattering-state contribution depend on the cutoff~$\zeta$ (diverge logarithmically). 
We find, both by an asymptotic analysis of the analytic formulas as well as numerically, that the divergent parts cancel each other. Therefore, the complete Josephson current per length is given by the limit of infinite cutoff:
\begin{align}
\frac{\iiJ(\phi)}{L_y} &= \lim_{\zeta \rightarrow \infty}
	\frac{\ii^\text{b}_\zeta + \ii^\text{sc}_\zeta}{L_y}~. \label{iiJtot}
\end{align}
In numerical analysis, the cutoff energy $\zeta$ has to be chosen such that sufficient convergence of the limit~\eqref{iiJtot} is achieved. 
Due to slower oscillations in the numerical integral, a larger value of $\zeta$ is needed for smaller values of $W/\xi$. 
As the numerical integration is less stable for larger $W$, multiple cutoff values are used.

\begin{figure}[t]
\hspace*{0.525cm}
\subfloat[SNS case]{\hspace{-0.55cm}\begin{overpic}[height=4.92cm]{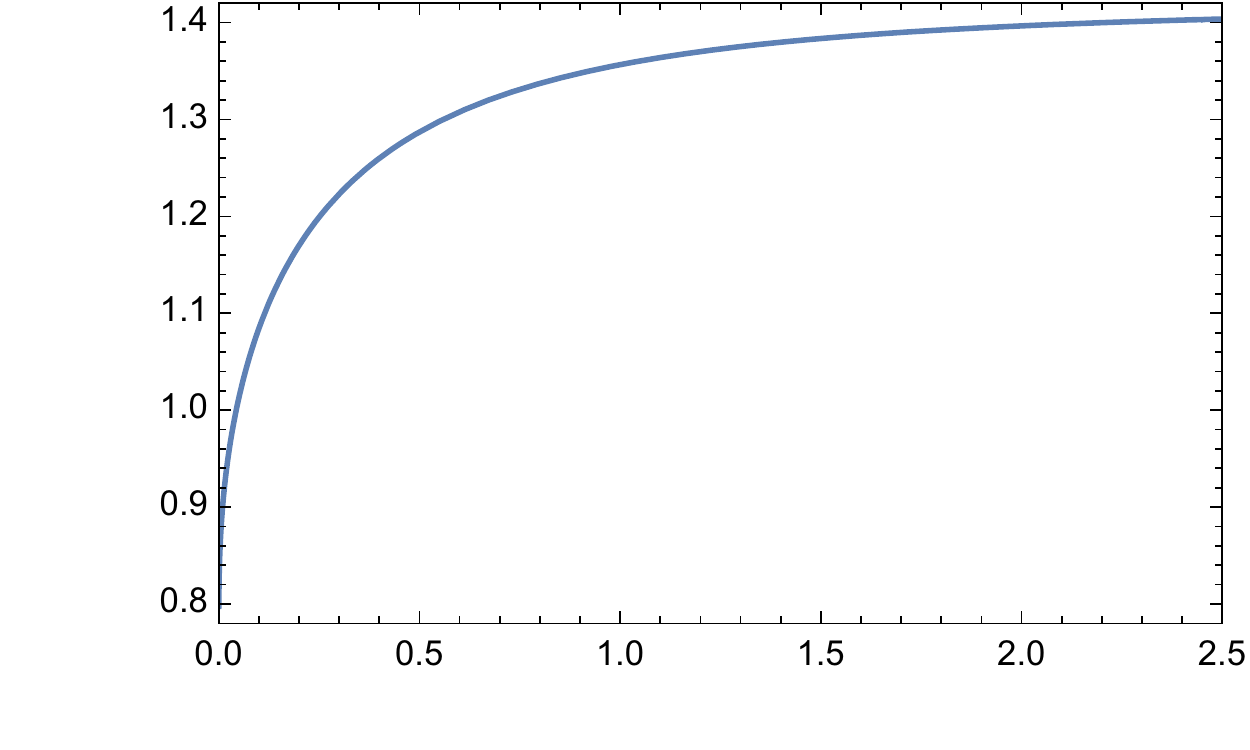}
	\put(53.5,2.5){\scalebox{1.3}{$\scriptstyle W / \:\!\xi$}}
	\put(-0.5,34){\scalebox{1.3}{$\frac{\iiJ(\phi)}{\iiJ'(0)\pI}$}}
	\put(30,35){\scalebox{1.3}{$\phi = \frac{3 \pi}{4}$}}
\end{overpic}\hspace{0.5cm}\label{J-SNSratio}}
\hspace{0.35cm}
\subfloat[SMS case]{\hspace{-0.55cm}\begin{overpic}[height=4.92cm]{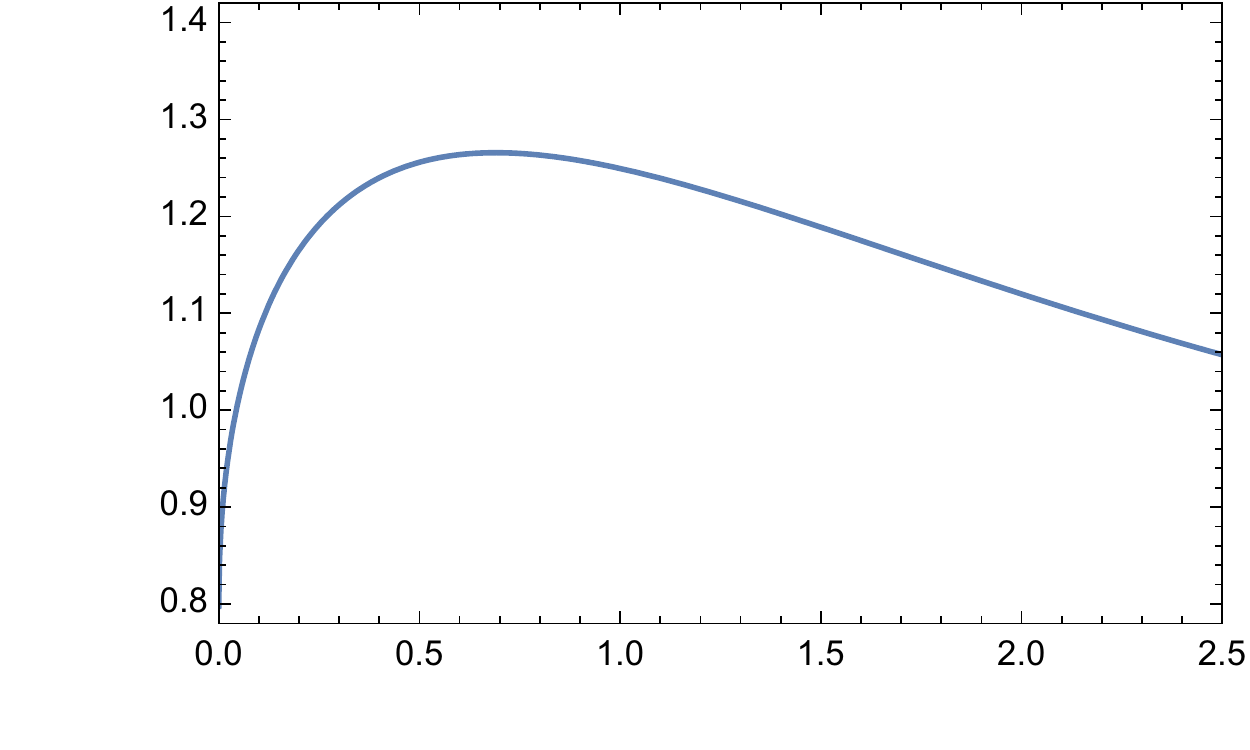}
	\put(53.5,2.5){\scalebox{1.3}{$\scriptstyle W / \:\!\xi$}}
	\put(-0.5,34){\scalebox{1.3}{$\frac{\iiJ(\phi)}{\iiJ'(0)\pI}$}}
	\put(30,35){\scalebox{1.3}{$\phi = \frac{3 \pi}{4}$}}
\end{overpic}\hspace{0.5cm}\label{J-SMSratio}}
\hspace*{-0.6cm}
\caption{The ratio between the current at $\phi = 3 \pi / 4$ and the slope at $\phi = 0$ quantifies the change of shape (cf.~Fig.~\ref{J-shapePlot}). 
In the SMS case~\subbf{J-SMSratio}, the ratio is non-monotonous as a function of the junction width~$W$. 
Plots for $\zeta/\Delta_0 = 1000$. 
}\label{J-ratioPlot}
\end{figure}

We start by presenting the current--phase relations ~$\iiJ(\phi)$ in Fig.~\ref{J-shapePlot} for several values of $W$. The curves are normalized to a uniform slope at $\phi = 0$, \ie, we plot $\iiJ(\phi) / \iiJ'(0) $, where $\iiJ'(0) \equiv \left. \frac{\pd\iiJ}{\pd\phi} \right|_{\phi = 0} $. This allows us to investigate the deviation of the current--phase relation from the sinusoidal one. In the SNS case, the deviation from the sinus shape increases with $W/\xi$ (Fig.~\ref{J-SNSchi}), while the behavior is non-monotonous in the magnetic case (Fig.~\ref{J-SMSchi}). 
As a numerical quantity characterizing the change of shape, the ratio between the current at $\phi = \frac{3 \pi}{4}$ and the slope $\iiJ'(0)$ is depicted for both cases as a function of $W$ in Fig.~\ref{J-ratioPlot}. 

\begin{figure}[t]
\hspace{0.0cm}
\subfloat[Decay for wide junctions]{\hspace{-0.5cm}\begin{overpic}[height=4.9cm]{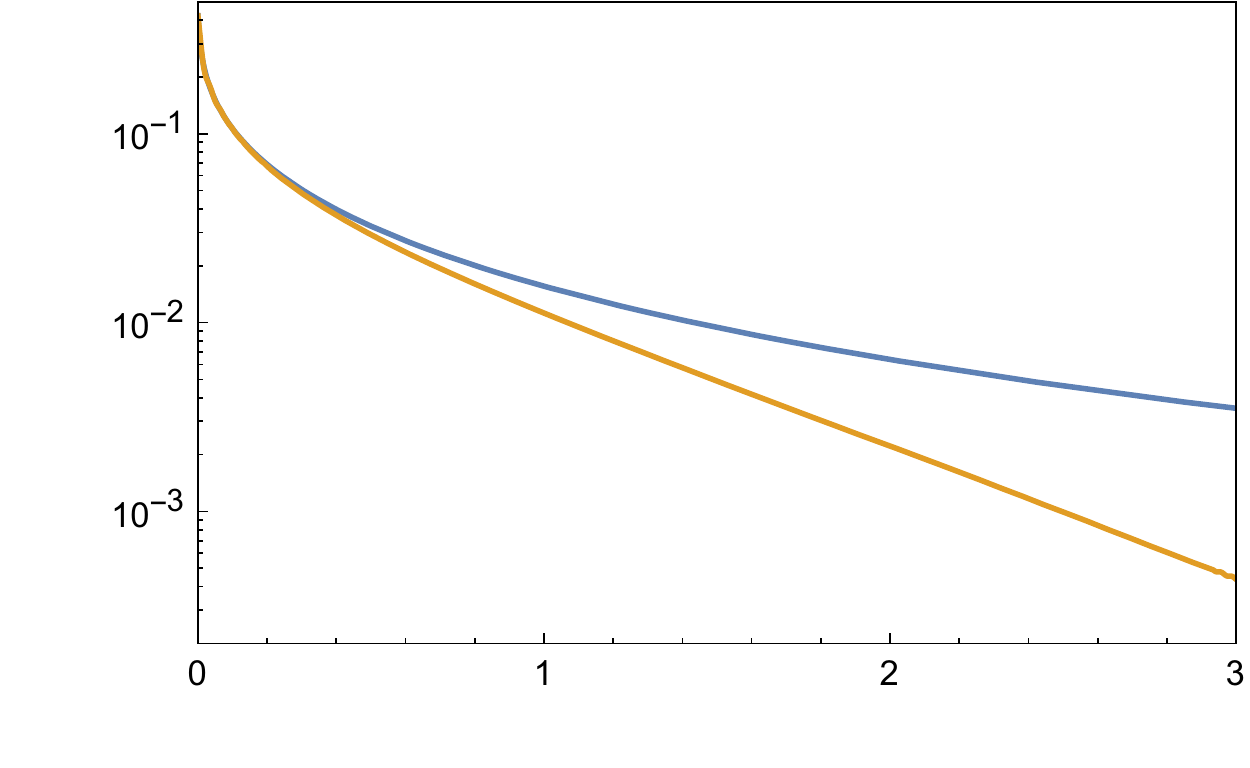}
	\put(53,2){\scalebox{1.3}{$\scriptstyle W / \:\!\xi$}}
	\put(-3.5,34){\scalebox{1.3}{$\frac{\iiJ'(0)}{\ii_0\pI}$}}
	\put(82,26.25){\scalebox{1.3}{$\scriptstyle\text{SNS}$}}
	\put(82,14.75){\scalebox{1.3}{$\scriptstyle\text{SMS}$}}
\end{overpic}\hspace{0.5cm}\label{J-linlog}}
\hspace{0.7cm}
\subfloat[double-logarithmic plot for the SNS case]{\hspace{-0.5cm}\begin{overpic}[height=4.9cm]{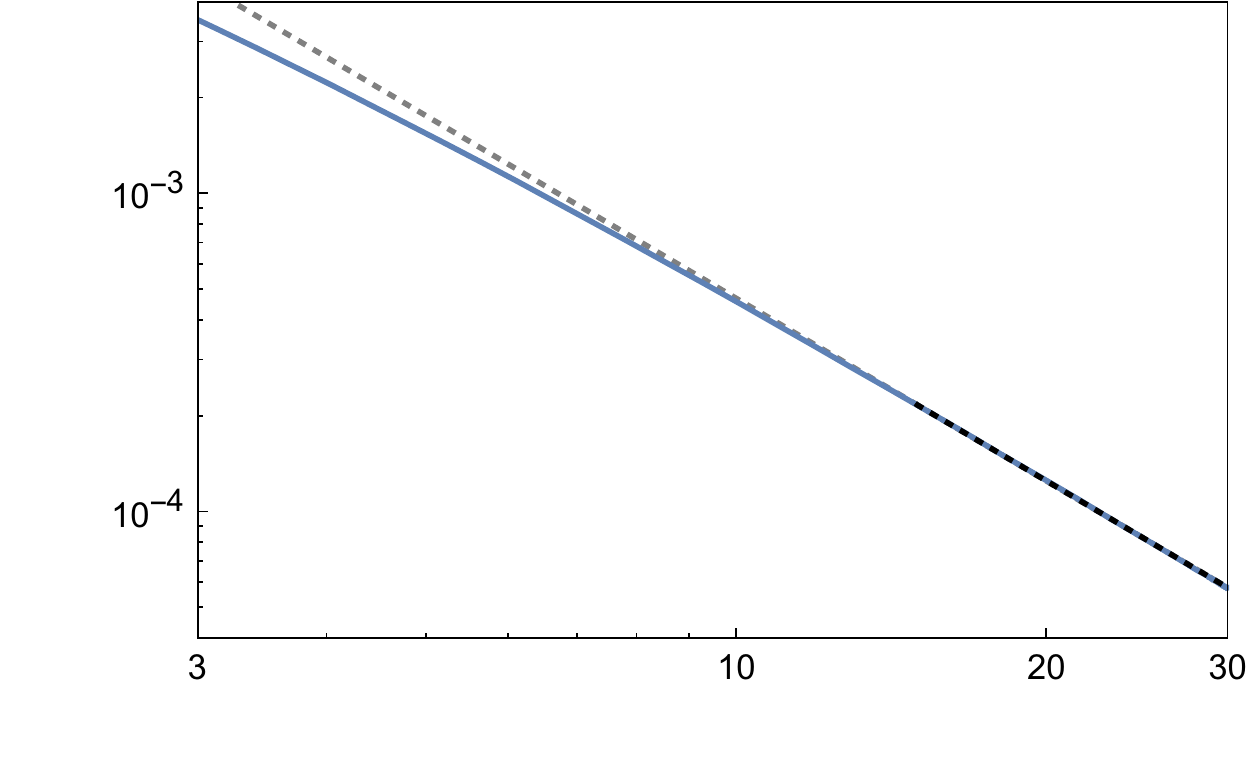}
	\put(54,2){\scalebox{1.3}{$\scriptstyle W / \:\!\xi$}}
	\put(-2.5,34){\scalebox{1.3}{$\frac{\iiJ'(0)}{\ii_0\pI}$}}
\end{overpic}\hspace{0.5cm}\label{J-SNSloglog}}
\hspace*{-0.75cm}

\hspace{0.98cm}
\subfloat[Growth for small junction width]{\hspace{-0.4cm}\begin{overpic}[height=4.93cm]{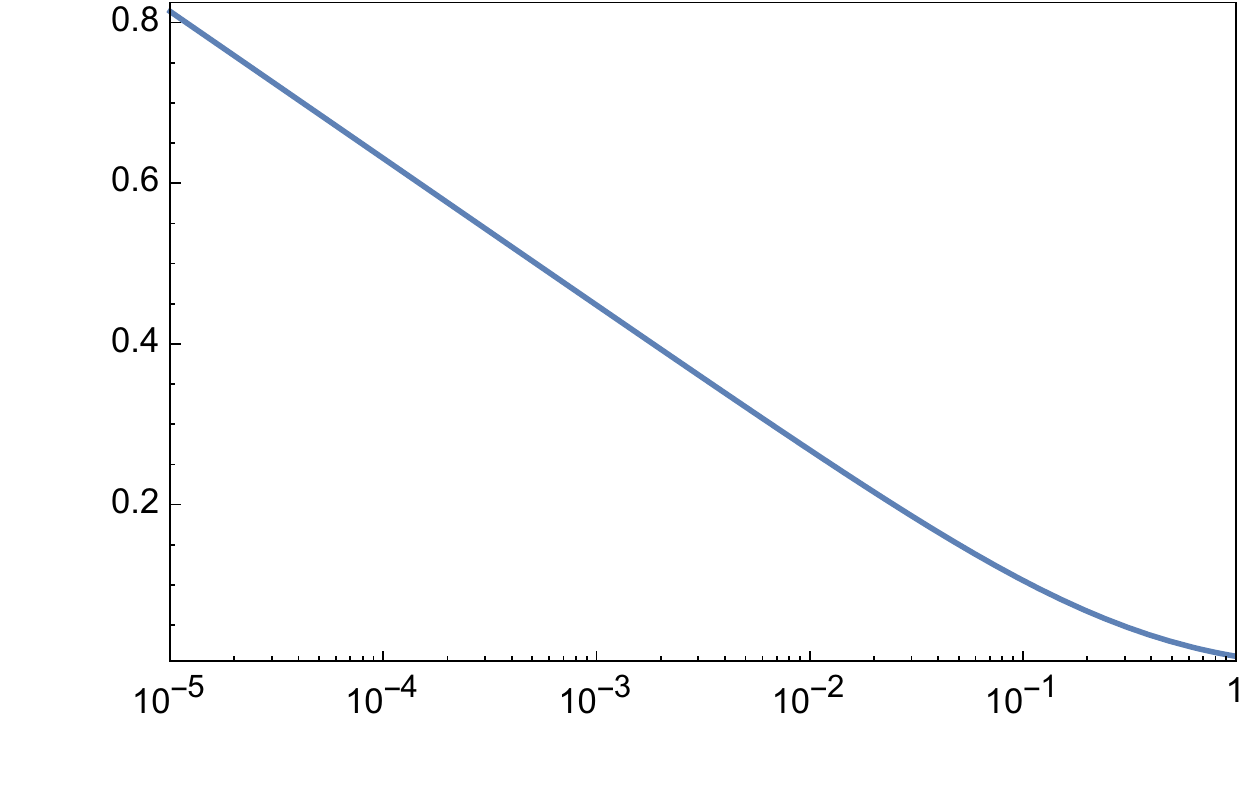}
	\put(52,2){\scalebox{1.3}{$\scriptstyle W / \:\!\xi$}}
	\put(-4.5,35){\scalebox{1.3}{$\frac{\iiJ'(0)}{\ii_0\pI}$}}
\end{overpic}\hspace{0.5cm}\label{J-loglin}}
\hfill~

\caption{The slope at $\phi = 0$ characterizes the magnitude of the current. 
For larger junction widths~$W$, this quantity is shown in the upper panels (both plots for $\zeta / \Delta_0 = 100$). 
\subbf{J-linlog} In the SMS case (orange curve), the current decays exponentially as a function of $W$, while the decay is slower without a magnetic gap. 
\subbf{J-SNSloglog} The width dependence looks approximately like a power law, $\iiJ \sim W^{-\alpha}$, for larger values of $W$ in the SNS~case. 
The power-law exponent~$\alpha$, which slightly increases with $W$, is given, \eg, by $\alpha \approx 1.9$ for $15 \le W/\xi \le 30$, represented by the~dotted~lines. 
\subbf{J-loglin} In the limit $W \rightarrow 0$, the current diverges logarithmically in both cases. 
}\label{J-logPlot}
\end{figure}
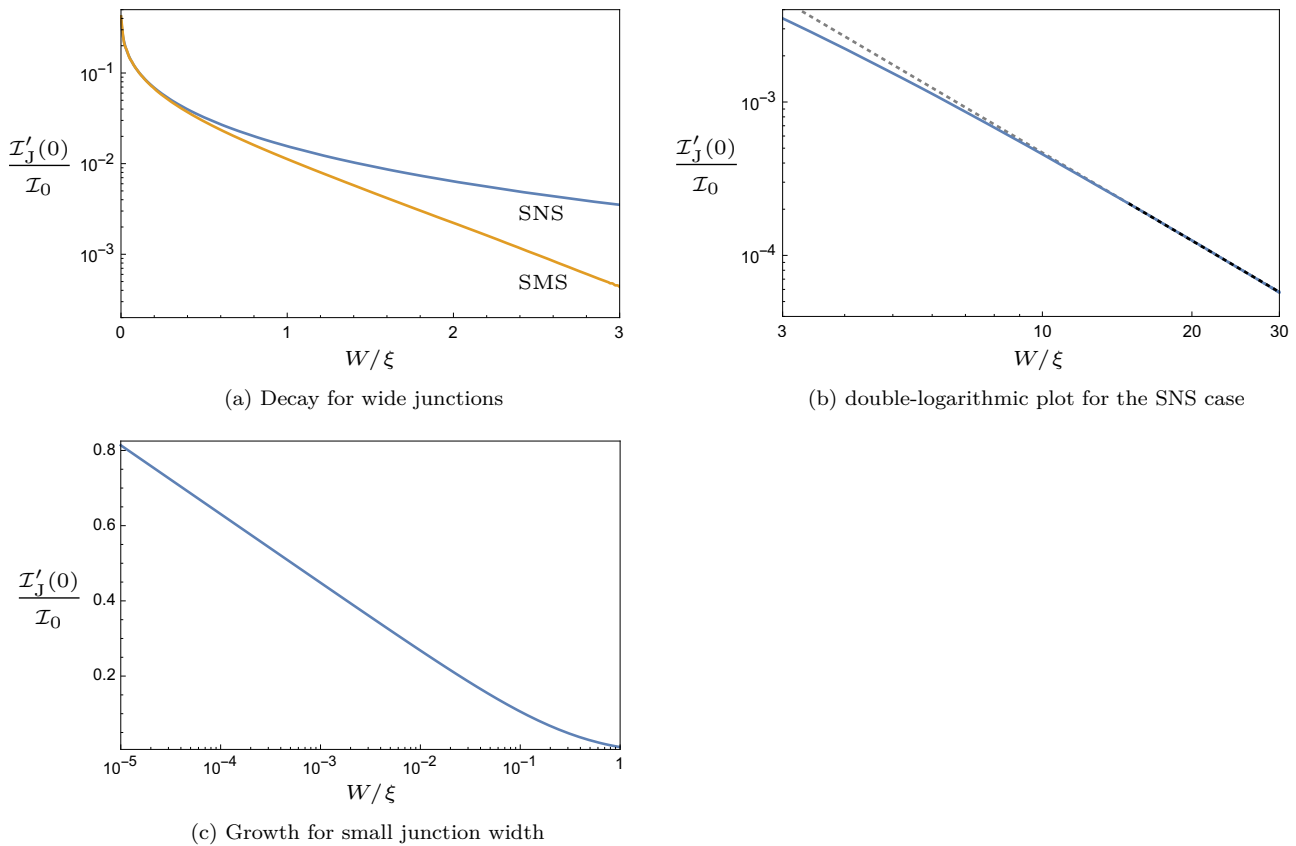

As a characteristic quantity for the magnitude of the current, we examine the derivative $\iiJ'(0)$ as a function of the junction width~$W$. 
The difference between non-magnetic and magnetic case is clearly visible in logarithmic plots: 
The current decays exponentially with $W$ due to the magnetic gap in the SMS case (Fig.~\ref{J-linlog}), whereas in the SNS system, an approximate power-law dependence is apparent from the double-logarithmic plot in Fig.~\ref{J-SNSloglog}. 
[In the SNS case, the contribution of Andreev bands and the one of scattering states individually show an oscillating behaviour, including sign changes. 
The simpler decay results only from the sum of both contributions, which is illustrated in the next section.] 

There is no difference between the SNS ($M_0=0$) and the SMS ($M_0=\Delta_0$) cases in the limit $W \ll \xi$ 
((Fig.~\ref{J-linlog})). However, in the numerical calculations, we have to treat the cutoff energy~$\zeta$ more carefully as it needs to be larger for narrow junctions. 
Analytical approximation of the integration~\eqref{scSNSii} over the expression~\eqref{IkqSMS}---we use the SMS expressions for simplicity---yields for the scattering modes a dominant contribution proportional to the logarithm of $W \, \zeta$ in the limit of small width~$W$:
$\ii^\text{sc}_\zeta \propto \ii_0 \, \ln\lft( \frac{W \, \zeta}{\xi \, \Delta_0} \right)$ for $W \ll \xi$ and  $\frac{W \, \zeta}{\xi \, \Delta_0} \gg 1$. As the logarithmic growth in the parameter~$\zeta$ cancels with the same divergence in the Andreev-band contribution~\eqref{IbSMS}, we obtain for the magnitude of the total current in the narrow-junction limit $W \ll \xi$
\begin{align}
\iiJ'(0) \propto \ii_0 \, \ln(\xi / W)\ .
\end{align}
This behavior is supported by a numerical result presented in Fig.~\ref{J-loglin}.

\section{Relative contributions of Andreev bands and scattering states}

\begin{figure}[h]
\hspace*{0.1cm}
\subfloat[SNS case]{\hspace{-0.5cm}\begin{overpic}[height=5cm]{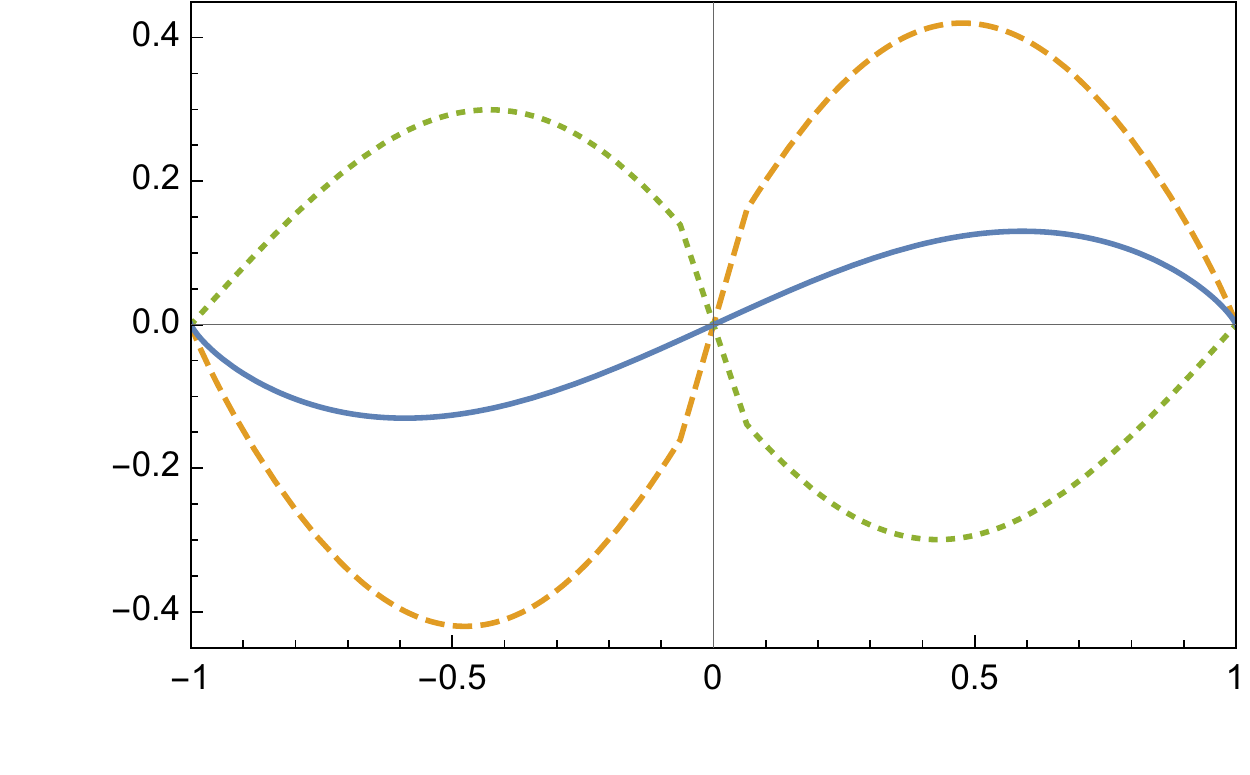}
	\put(53,2){\scalebox{1.3}{$\myfrac{\phi\,}{\;\!\pi}$}}
	\put(-2,35){\scalebox{1.3}{$\frac{\ii(\phi)}{\ii_0\pI}$}}
\end{overpic}\hspace{0.5cm}\label{J-SNS01}}
\hspace{1cm}
\subfloat[SMS case]{\hspace{-0.5cm}\begin{overpic}[height=5cm]{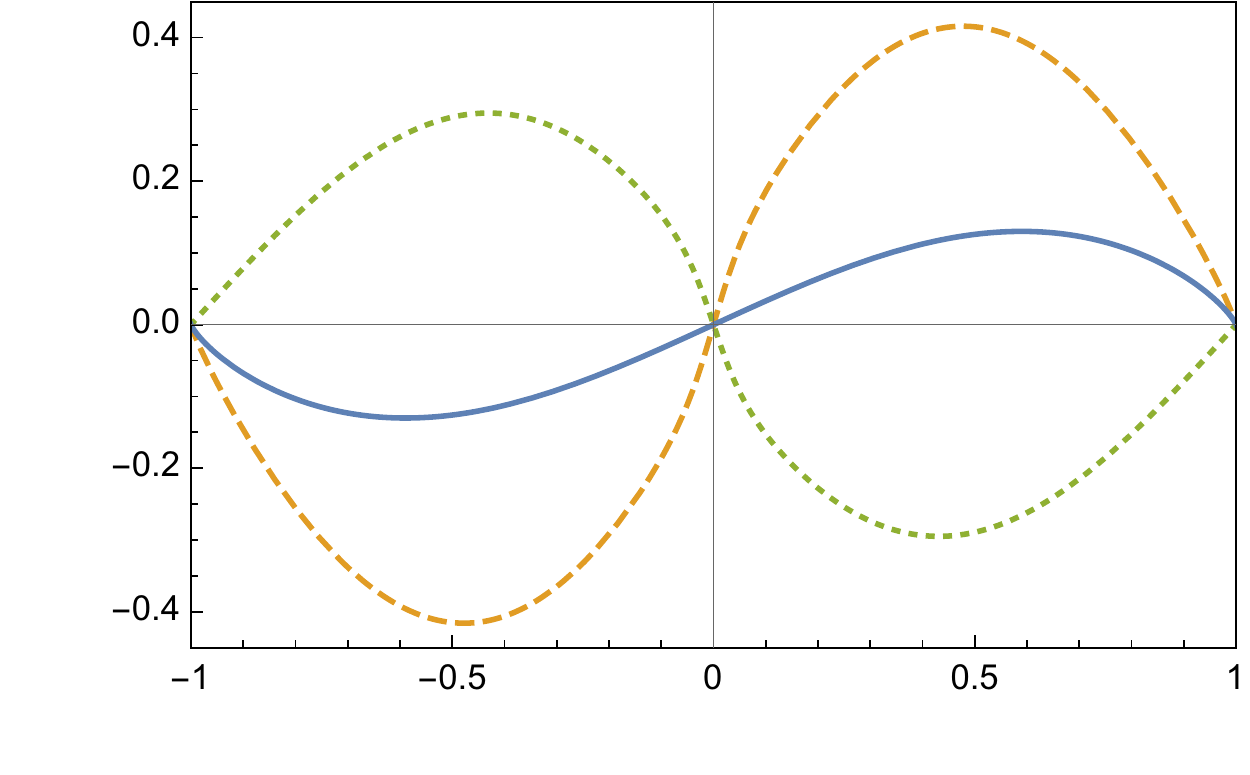}
	\put(53,2){\scalebox{1.3}{$\myfrac{\phi\,}{\;\!\pi}$}}
	\put(-2,35){\scalebox{1.3}{$\frac{\ii(\phi)}{\ii_0\pI}$}}
\end{overpic}\hspace{0.5cm}\label{J-SMS01}}
\hspace*{-0.7cm}
\caption{Current contributions of Andreev bands $\ii^\text{b}_\zeta$ (dashed in orange) and of scattering states $\ii^\text{sc}_\zeta$ (dotted in green) as well as the total Josephson current $\iiJ$ (blue line) for a narrow junction. 
\subbf{J-SNS01} Except for a small region around $\phi = 0$, the Andreev-state contribution results from a single band in the SNS case, too, which is therefore similar to~the magnetic case in~\subbf{J-SMS01}. 
Plots for $W / \xi = 0.1$, $\zeta / \Delta_0 = 100$.
}\label{J-multiB01}
\end{figure}

\begin{figure}[h]
\hspace*{0.05cm}
\subfloat[SNS case]{\hspace{-0.5cm}\begin{overpic}[height=5cm]{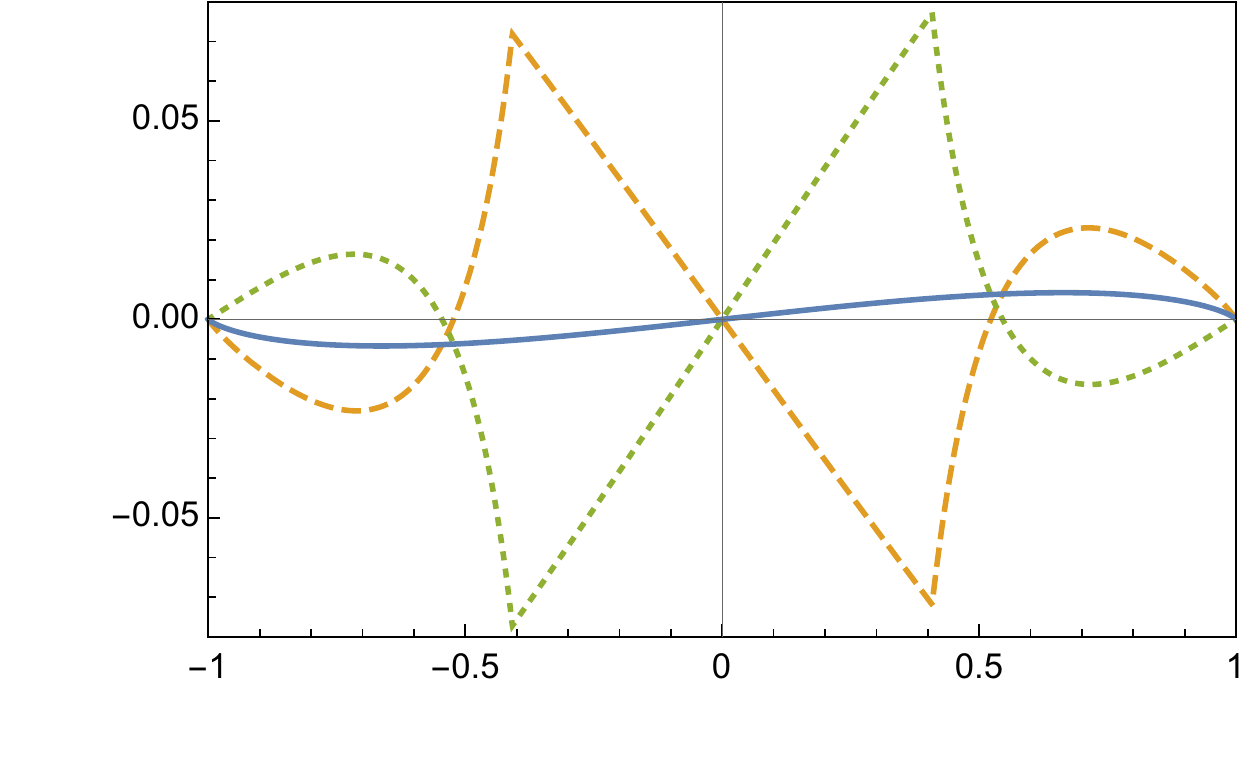}
	\put(53.8,2){\scalebox{1.3}{$\myfrac{\phi\,}{\;\!\pi}$}}
	\put(-1,34.5){\scalebox{1.3}{$\frac{\ii(\phi)}{\ii_0\pI}$}}
\end{overpic}\hspace{0.5cm}\label{J-SNS25}}
\hspace{0.8cm}
\subfloat[SMS case]{\hspace{-0.6cm}\begin{overpic}[height=5cm]{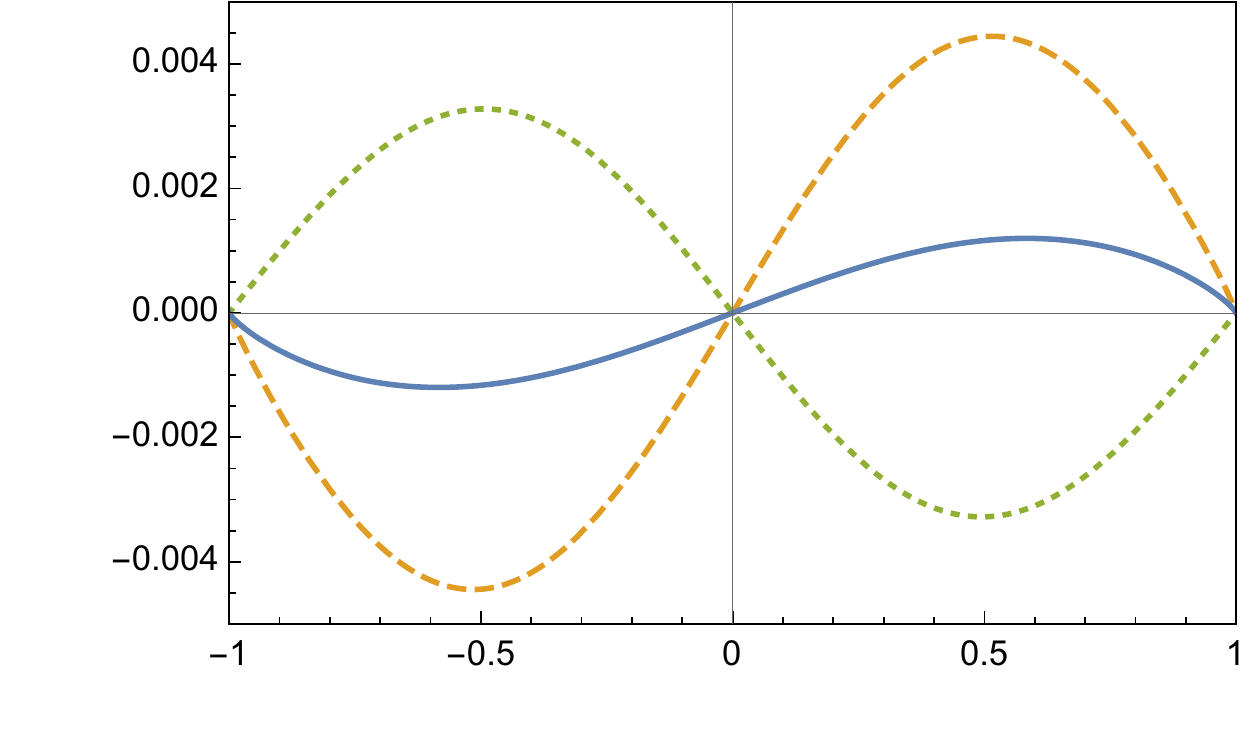}
	\put(54.75,2){\scalebox{1.3}{$\myfrac{\phi\,}{\;\!\pi}$}}
	\put(-1,34.1){\scalebox{1.3}{$\frac{\ii(\phi)}{\ii_0\pI}$}}
\end{overpic}\hspace{0.6cm}\label{J-SMS25}}
\hspace*{-0.7cm}
\caption{Current contributions of Andreev bands $\ii^\text{b}_\zeta$ (dashed in orange) and of scattering states $\ii^\text{sc}_\zeta$ (dotted in green) as well as the total Josephson current $\iiJ$ (blue line). 
\subbf{J-SNS25} For a wider junction, multiple modes (cf.~Fig.~\ref{J-mSNS25}) affect the Andreev-state contribution in the SNS case. 
However, kinks and oscillations due to the Andreev bands are perfectly compensated by the scattering-state contribution, yielding a smooth result. 
\subbf{J-SMS25} In the SMS case, the current decays faster as a function of junction width~$W$ (cf.~Fig.~\ref{J-linlog}), and the separate contributions are simpler, 
as there is always a single Andreev band, only. 
Plots for $W / \xi = 2.5$, $\zeta / \Delta_0 = 100$.
}\label{J-multiB25}
\end{figure}

It is instructive to plot the contributions of the Andreev bands and those of the scattering states 
separately. Since both are logarithmically divergent, a certain cutoff $\zeta / \Delta_0$ must be 
assumed. The results are shown in Fig.~\ref{J-multiB01} and Fig.~\ref{J-multiB25}.

We observe two very different behaviors in the limit of a wide junction ($W\gg \xi$) in the SNS case (Fig.~\ref{J-SNS25}) 
and in all other regimes, \eg, for narrow SNS and SMS junctions (Fig.~\ref{J-multiB01}) and for a wide SMS junction (Fig.~\ref{J-SMS25}). In the latter three regimes, the contribution of the Andreev states is (mostly) dominated by a single band. 
(However, the kinks in Fig.~\ref{J-SNS01} are due to the merge of further Andreev modes with the 
continuum.) The contribution of the scattering states has an opposite sign and removes the logarithmic dependence on the 
cutoff. In the regime of Fig.~\ref{J-SNS25}, competing contributions of multiple Andreev bands are evident, giving rise 
to kinks, oscillations and, even, a $\pi$-junction behavior. Yet, the scattering states compensate all these features, producing a regular \emph{sine}-like current--phase relation.

\section{Finite temperature}

In this section, we consider the regime of low but non-zero temperatures, $0 < k_\text{B} \, T \ll \Delta_0$. In this regime, the thermal factor can be neglected in the scattering-state contributions to the Josephson current but not in those due to the Andreev bands. 
The integration over Andreev bands, Eq.~\eqref{bSNSii}, has to be carried out 
numerically for $T > 0$. In the following, we focus on the SNS case because it shows more pronounced finite-temperature effects on the Josephson current $\iiJ(\phi)$. 

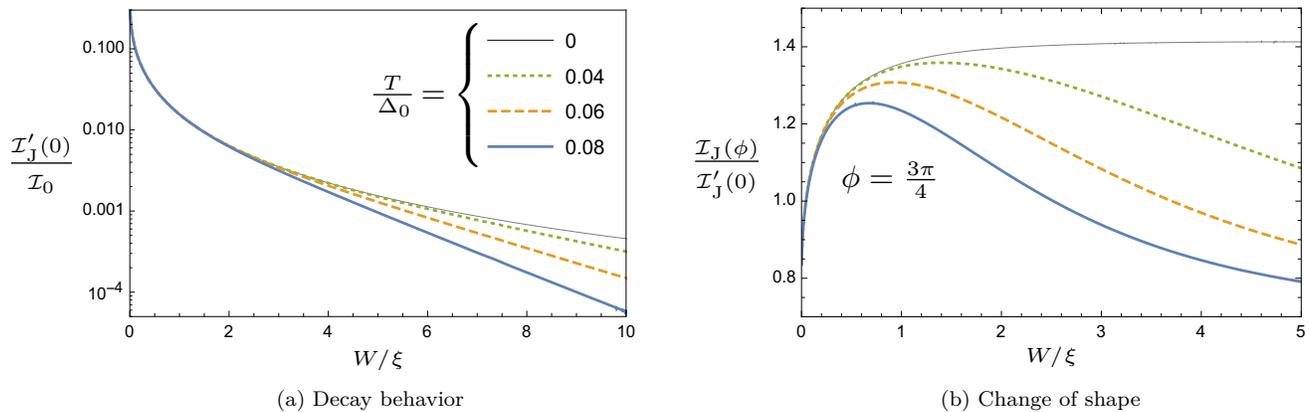
\begin{figure}[t]
\hspace{0.0cm}
\subfloat[Decay behavior]{\hspace{-0.5cm}\begin{overpic}[height=4.9cm]{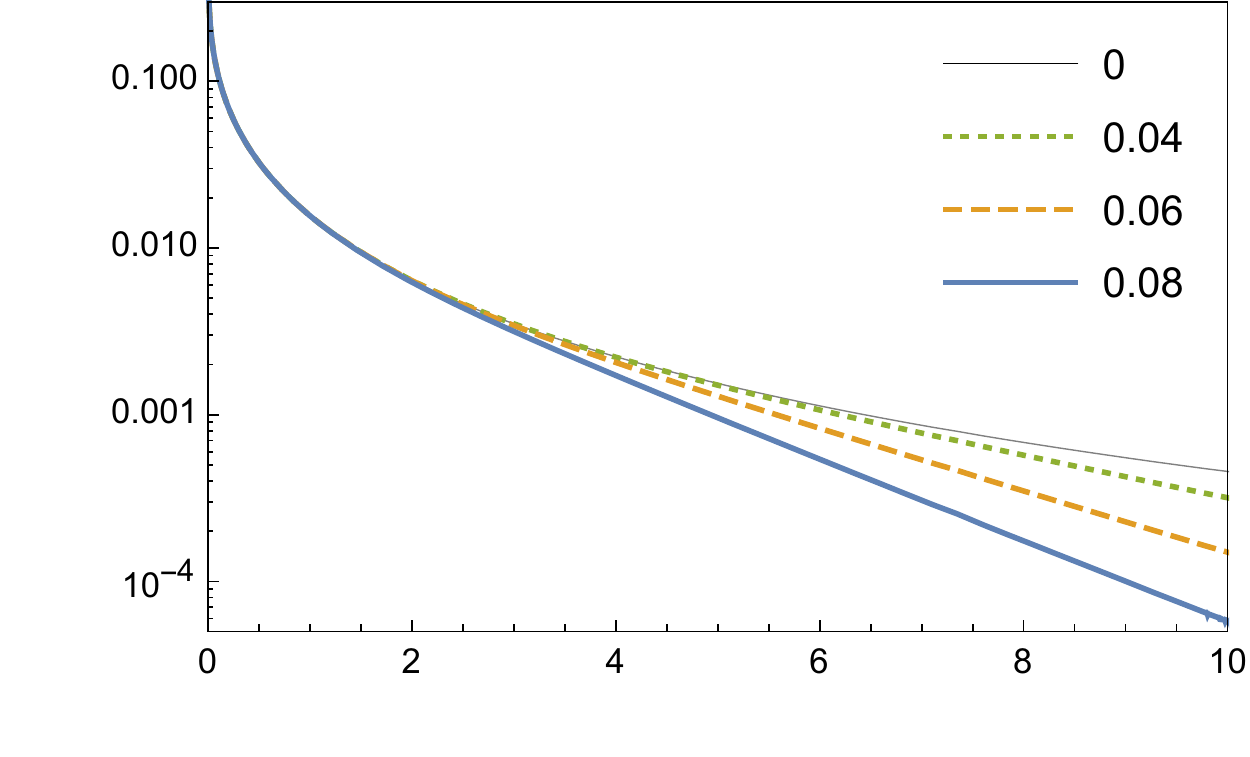}
	\put(53.25,2){\scalebox{1.3}{$\scriptstyle W / \:\!\xi$}}
	\put(-3.5,34){\scalebox{1.3}{$\frac{\iiJ'(0)}{\ii_0\pI}$}}
	\put(56,45.25){\scalebox{1.3}{$\frac{T}{\,\Delta_0} \, {=} \: \bigger\{$}}
\end{overpic}\hspace{0.5cm}\label{JT-SNSlinlog}}
\hspace{0.7cm}
\subfloat[Change of shape]{\hspace{-0.55cm}\begin{overpic}[height=4.92cm]{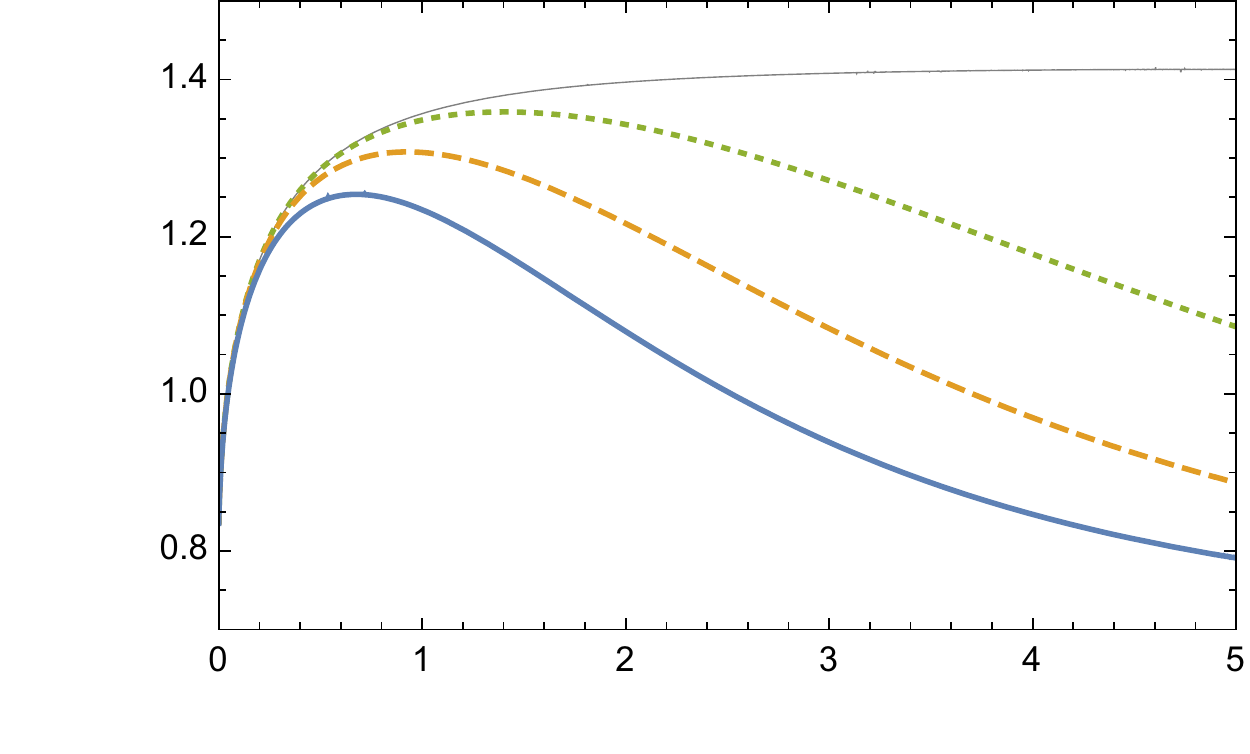}
	\put(54,2.5){\scalebox{1.3}{$\scriptstyle W / \:\!\xi$}}
	\put(-0.5,34){\scalebox{1.3}{$\frac{\iiJ(\phi)}{\iiJ'(0)\pI}$}}
	\put(24,31){\scalebox{1.3}{$\phi = \frac{3 \pi}{4}$}}
\end{overpic}\hspace{0.5cm}\label{JT-SNSratio}}
\hspace*{-0.7cm}
\caption{Josephson current in the SNS case for non-zero temperatures. 
\subbf{JT-SNSlinlog} The slope at $\phi = 0$ characterizes the magnitude of the current (see also Fig.~\ref{J-logPlot}). At non-zero temperatures, the dependence of the slope
on the junction width~$W$ crosses over to an exponential decay. 
\subbf{JT-SNSratio} As a measure for the deviation of the current--phase relation from the sinusoidal one, we use the ratio between the current at $\phi = 3 \pi / 4$ and the slope at $\phi = 0$, like in Fig.~\ref{J-ratioPlot}. 
At~non-zero temperatures~$T$, the SNS Josephson current--phase relation re-approaches the sinusoidal shape (ratio~$\sim 0.7$) for larger junction widths~$W$. 
Plots for $\zeta / \Delta_0 = 100$. 
}\label{JT-SNS}
\end{figure}

The thermal factor is relevant for an Andreev band if the lower band edge, the effective mass~$m$ (Fig.~\ref{J-mSNS}), is of the order of the thermal energy~$k_\text{B} \, T$ or smaller. 
Hence, for small junction widths $W \ll \xi$, the Josephson current is only slightly changed around the zero of $m$ at phase difference~$\phi = \pi$. 
For increasing junction width, the lowest Andreev-band edge at $\phi = 0$ drops below $k_\text{B} \, T$ at some point. 
This leads to a crossover of the phase derivative $\iiJ'(0)$ from the approximate power-law behavior in the zero-temperature SNS case to an exponential decay as a function of junction width (see~Fig.~\ref{JT-SNSlinlog}). 

Unlike in the $T = 0$ analysis of Sec.~\ref{zero}, the deviation of the current--phase relation from the sinusoidal shape shows a non-monotonous behavior in the SNS case, too. 
In order to quantify the change of shape, the ratio between the current at $\phi = \frac{3 \pi}{4}$ and the slope $\iiJ'(0)$ is depicted in Fig.~\ref{JT-SNSratio}. 
The plot shows that at finite temperature, the current--phase relation is approximately sinusoidal not only for very narrow junctions (cf.~Fig.~\ref{J-SNSchi}), but also for large values of $W$, dependent on the temperature~$T$. 

\section{Discussion}\label{sec:Discussion}

The limit $\mu = 0$ (Dirac point) of a long, translation-invariant Josephson junction on a TI surface has some peculiar properties. 
In particular, Andreev bands of states localized around the junction are not restricted to energies below the continuum, which results in their logarithmically divergent contribution to the Josephson current. To compensate for that, the same divergence exists in the scattering-state contribution. 
As the translation-invariant direction trivially decouples from the physics across the junction, the spectra of effective masses~$m$ for the 1D Andreev \emph{bands} in our model resemble the energies of \emph{zero-dimensional} Andreev states obtained by Fu and Kane\cite{FuKanePRB2009} for a Josephson junction in an effectively one-dimensional topological superconductor. 

For wide junctions, $W \ge \xi$, the magnetic and non-magnetic cases differ significantly. 
In the SNS case, the current as a function of $W$ behaves in a way similar to a non-topological superconductor--metal--superconductor junction, including oscillations in the contributions of Andreev bands and of scattering modes which cancel each other \cite{mqp, jcn}. The magnetic case involves a simpler exponential decay of both contributions, but the shape of the current--phase relation changes in a non-monotonous and hence more complicated way. 

In contrast, in the narrow-junction limit, $W\ll \xi$, a finite magnetic gap does not make much of a difference. 
In this limit, it is instructive to compare the width dependence of the Josephson current in our case to the analysis of Titov and Beenakker\cite{TitovBeenakkerPRB2006} for the critical current $I_\text{c}$ of an SNS Josephson junction at the Dirac point of graphene. (The Dirac point was assumed in \Rcite{TitovBeenakkerPRB2006} only for the normal stripe of graphene between the superconductors, whereas the chemical potential of graphene is large in the superconducting leads.) The normal resistance of a graphene stripe with sizes $W$ and $L_y$ can be estimated at the Dirac point 
as $R_N \sim (\hbar/e^2) \, W/L_y$, so from the Ambegaokar--Baratoff relation $I_\text{c} \, R_N \sim \Delta_0/e$,
one can expect 
\begin{align}
I_\text{c} &\sim \frac{e \, \Delta_0}{\hbar} \, \frac{L_y}{W}
\qquad \text{for} \quad W \ll \xi~. \nonumber
\end{align}
This behavior was indeed obtained in \Rcite{TitovBeenakkerPRB2006}. 

In our model, however, the chemical potential vanishes in the whole system, $\mu \equiv 0$, so a normal-state resistance cannot be determined: 
Without superconductivity, at least in the SNS case, there is no junction at all. 
Using the definitions of Eq.~\eqref{defs} and the $W$ dependence discussed in Sec.~\ref{zero}, we obtain
\begin{align}
I_\text{c} &\sim \frac{e \, \Delta_0}{\hbar} \, \frac{L_y}{\xi} \, \ln\lft( \frac{\xi}{W} \right)
\qquad \text{for} \quad W \ll \xi~, \label{Ic0}
\end{align}
instead. 
Here, the contribution of the continuous spectrum of scattering states is essential and cannot be neglected even in this limit. 
This means that the chemical potential being at the Dirac point in the superconducting regions, too, leads to a qualitatively different behaviour for narrow junctions. 

The effects of finite, low temperatures, $0 < k_\text{B} \, T \ll \Delta_0$, are most prominent for wide junctions, in particular in the SNS case. 
As the widths $W$ grows, the temperature induces both an exponential decay of the current magnitude and a reentrant trend towards a sinusoidal shape of the current--phase relation. Therefore, the qualitative difference between the SNS and the SMS cases is smaller at finite temperatures. 

In general, the compensation of two logarithmic divergences is only possible if the linear spectrum of the topological insulator holds for an unlimited range of energy. 
The integration range that is effectively required to observe the compensation rises up to infinity when the junction width~$W$ is decreased to zero. 
Therefore, the logarithmic divergence of the critical current, Eq.~\eqref{Ic0}, as a function of $W$ will be cut off at some point for a more realistic spectrum. 
Moreover, for experimental realizations, an intermediate regime of the magnetic gap, $0 < M_0 < \Delta_0$, is likely to be relevant. 
There, it is reasonable to assume that the behavior of the Josephson junction will fall somewhere between the SNS case and the $M_0 = \Delta_0$ regime we have analyzed. 
In the limit $M_0 \gg \Delta_0$, on the other hand, the large magnetic gap will lead to a strong suppression of the Josephson current. This increasing $M_0$ should have a similar effect to increasing $W$ for a fixed $M_0$.

While the analytical and numerical calculations presented here rely on the strict assumption $\mu = 0$, we do not expect significant changes in the regime of a small chemical potential, $|\mu| \ll \Delta_0$: 
A perturbative analysis within the state space of an Andreev band and its negative-energy counterpart up to first order in the chemical-potential contribution to the Hamiltonian, $-\mu \, \tau_z$, yields vanishing energy and state corrections. 
This indicates only a weak influence on the behavior of the system at low energies. 
As this perturbation preserves the translation invariance in $y$~direction, a small chemical potential should not cause large changes via contributions of higher-energy states, either.

\section{Conclusions}

In this paper, we have studied the current--phase relation of long topological Josephson junctions.
We have considered two simple cases where it was possible to fully take into account both the 
contribution of the Andreev bound states (Andreev bands) and that of the continuum of scattering states. 
Both contributions turn out to be logarithmically divergent, but these divergencies compensate 
each other in the total result. Moreover, the contribution of the Andreev bands can show very unusual 
behavior, \eg, negative slope of the Josephson current at zero phase bias ($\pi$-junction) as well as 
kinks and zero-crossings at phase biases different from $0$ and $\pi$. Yet, all these features are 
completely eliminated by the corresponding contribution of the scattering states. The current--phase relation 
remains close to sinusoidal with some deviations towards the saw-tooth shape. This shows that 
taking into account both contributions is crucial. We have also analyzed the Josephson current in the limit of a very thin (narrow) junction and discussed the 
applicability of the Ambegaokar--Baratoff relation.

\acknowledgments 
This research was supported by DFG Grant SH 81/4-1 and by joint 
DFG-RFBR Grant (SH 81/6-1 (DFG), 20-52-12034 (RFBR)). We are grateful to 
Yu. Makhlin and M. Titov for encouraging discussions.

\appendix

\section{Eigenmodes and supercurrent in a thermal state}\label{appSuperEigen}

We provide here the standard derivation of the Josephson current in order to relate it to the eigenmodes 
of the single particle Hamiltonian $h$ introduced in Eq.~\eqref{NambuH}. These eigenmodes satisfy 
$h \,\Xi_j = \epsilon_j \,\Xi_j$, where $\Xi_j(x,y)= [\xi_{1j}(x,y), \xi_{2j}(x,y), \xi_{3j}(x,y), \xi_{4j}(x,y)]^\T$ is a 4-component spinor wave function. Due to the particle--hole (charge-conjugation) symmetry, every positive energy solution $\Xi_j$, $\epsilon_j > 0$ has a negative energy counterpart, $\Xi_{-j} = \tau_y \sigma_y \, \Xi_j^*$ with $\epsilon_{-j}=-\epsilon_j$.
The second-quantized Hamiltonian $H$ of Eq.~\eqref{NambuH} is then diagonalized as 
\begin{align}
H &=  \sum_{\epsilon_j>0} \epsilon_j \, \left[\alpha_j^\dag \, \alpha_j^\0 -\frac{1}{2}\right]\label{eigen.diagH}~,
\intertext{where the sum runs over the positive energies only (for the continuous part of the spectrum, the sums are to be taken as integrals, of course) and the quasiparticle creation operator reads}
\alpha_j^\dag &= \int\! \td x \, \td y \, \Psi^\dag(x, y) \, \Xi_j(x, y)~.
\end{align}
The Josephson current is the derivative of the free energy $F = -\beta^{-1} \, \ln(\Z)$ with respect to the phase difference~$\phi$, $\iiJ = \frac{2 \, e}{\hbar} \, \frac{\pd F}{\pd\phi}$. For the fermionic partition function one obtains 
\begin{align}\label{PartitionZ}
\ln(\Z) = \sum_{\epsilon_j>0} \, \ln\lft(2 \, \cosh\lft[\frac{\beta \, \epsilon_j}{2}\right]\right)~,
\end{align}
which gives
\begin{align}
\iiJ = \frac{2 \, e}{\hbar} \, \frac{\pd F}{\pd\phi} = -\frac{1}{2} \sum_{\epsilon_j>0} \left[\frac{2 \, e}{\hbar} \, \frac{\pd \epsilon_j}{\pd\phi}\right] \tanh\lft( \frac{\beta \, \epsilon_j}{2} \right) = -\frac{1}{2} \sum_{\epsilon_j>0} I_j \, \tanh\lft( \frac{\beta \, \epsilon_j}{2} \right)
~, \label{super.iiJ}
\end{align}
where we have introduced a partial current $I_j \equiv \frac{2 \, e}{\hbar} \, \frac{\pd \epsilon_j}{\pd\phi}$ corresponding to
the single-particle state $j$. 

For the Andreev bands, Eq.~\eqref{super.iiJ} is the most convenient way to calculate the Josephson current. 
Let us consider the continuum of scattering states. There are two alternative ways to proceed: One 
can express $I_j$ as a function of the relevant quantum numbers of the state $j$. (In our case, the two 
momenta $q$ and $k$ are quantum numbers and for each pair $(q,k)$, 
there are four different scattering states.) Then one 
converts \eqref{super.iiJ} into an integral. This way, it is crucial to normalize the continuum states properly. 

The second way is to write the contribution of the scattering states to the partition function \eqref{PartitionZ} as
\begin{align}
\ln(\Z^\text{sc}) =  \int_{0}^{\zeta} \td\epsilon\, \rho^\text{sc}(\epsilon) \, \ln\lft(2\cosh\lft[\frac{\beta \, \epsilon}{2}\right]\right)~,
\end{align}
where $\zeta$ is the ultraviolet cutoff and $\rho^\text{sc}(\epsilon)$ is the density of states for the scattering states. 
In this integral, only $\rho^\text{sc}(\epsilon)$ can depend on the phase bias, thus 
\begin{align}
\iiJ = \frac{2 \, e}{\hbar} \, \frac{\pd F}{\pd\phi} =
- \frac{2 \, e}{\hbar \, \beta} \, \int_{0}^{\zeta} \td\epsilon\, \left[\frac{\pd\rho^\text{sc}(\epsilon)}{\pd \phi}\right]\, \ln\lft(2 \, \cosh\lft[\frac{\beta \, \epsilon}{2}\right]\right)~.
\end{align}
One could wonder how the 2D density of the scattering states $\rho^\text{sc}(\epsilon)$, 
which is an extensive thermodynamic quantity, can depend on the 
phase bias across the junction. Indeed, the spectrum of the scattering sates 
$\epsilon(k,q) = \sqrt{\Delta_0^2 + v^2 \, (k^2 + q^2)}$ is independent of $\phi$.
The solution is that, in general, the density of states~$\rho^\text{sc}(\epsilon)$ consists of both extensive and subextensive contributions. For a two-dimensional system of area $L^2$, \eg, the asymptotic behavior 
in the limit $L \rightarrow \infty$ is given by
\begin{align}
\rho^\text{sc} = \rho_2 \, L^2 +  \rho_1 \, L  + \dots ~.
\end{align}
It is the leading contribution $\rho_2$ which is independent of $\phi$, 
whereas the sub-leading contributions can depend on $\phi$. Since 
$\iiJ \propto L$, it is the $\rho_1$ contribution which is responsible for 
the Josephson current.
A more detailed analysis has been carried out for similar scaling behaviour of flux-dependent persistent currents in \Rcite{pcs}. 

In this paper, we pursue the first strategy, based on \eqref{super.iiJ}, and evaluate the partial current $I_j$
for the scattering states. Using $\epsilon_j=\left< \Xi_j \right| h \left| \Xi_j \right>$, we get 
\begin{align}
I_j = \frac{2 \, e}{\hbar} \, \frac{\pd \epsilon_j}{\pd\phi} = 
\frac{2 \, e}{\hbar} \left< \Xi_j \right| \frac{\pd h}{\pd \phi} \left| \Xi_j \right>~. \label{eigen.IE}
\end{align}
First, we apply the gauge transformation $\left| \Xi_j \right> \rightarrow  \e^{\I \gauge(x)\tau_z/2} \, \left| \Xi_j \right>$
in order to shift the phase dependence in the Hamiltonian from the order parameters in the superconducting areas to the middle region. The phase $\gauge(x)$ depends on the coordinate $x$; it has to satisfy $\gauge(x) = \phiL$ for $x < -W/2$ and $\gauge(x) = \phiR$ for $x > W/2$. In the middle region, the choice of gauge is not important and does not need to be fixed yet. The corresponding transformation of the Hamiltonian reads
\begin{align}
h &\rightarrow v \, \pvec \cdot \sigvec \, \tau_z + M(x) \, \sigma_z
	+ \Delta_0 \, \tau_x - \frac{\hbar \, v}{2} \, \sigma_x \, \Dgauge~.
\end{align}
Thus, in the new gauge, we obtain
\begin{align}
\frac{\pd h}{\pd \phi} = -\frac{\hbar \, v}{2}\,\sigma_x \,\DDgauge
\end{align}
and
\begin{align}\label{Ijxixi}
I_j = 
\frac{2 \, e}{\hbar} \left< \Xi_j \right| \frac{\pd h}{\pd \phi} \left| \Xi_j \right>=
-e \, v \int\! \td x \, \td y \, \DDgauge
	\, \left[ \xi_{1,j}^* \, \xi_{2,j}^\0 + \xi_{2,j}^* \, \xi_{1,j}^\0 + \xi_{3,j}^* \, \xi_{4,j}^\0 + \xi_{4,j}^* \, \xi_{3,j}^\0  \right]~.
\end{align}

What remains to be fixed is the gauge $\gauge(x)$ in the region between the superconductors. 
As, in fact, the rest of the integrand in \eqref{Ijxixi} does not depend on position in the interval $-W/2 \le x \le W/2$, it does not matter how exactly $\gauge$ is defined there. 
A simple choice is a jump at $x = 0$,
\begin{align}
\Dgauge = -\underbrace{(\phiL - \phiR)}_{\textstyle \phi} \, \delta(x) 
\quad \Rightarrow \quad \DDgauge = -\delta(x)~,
\end{align}
yielding
\begin{align}
I_j &= e \, v \int\! \td y \, \bigl[ \xi_{1,j}^* \, \xi_{2,j}^\0 + \xi_{2,j}^* \, \xi_{1,j}^\0
	+ \xi_{3,j}^* \, \xi_{4,j}^\0 + \xi_{4,j}^* \, \xi_{3,j}^\0 \bigr]_{x=0}~. \label{eigen.Ixi}
\end{align}
Finally, we note that the expression \eqref{eigen.Ixi} is invariant with respect to the gauge transformation 
$\left| \Xi_j \right> \rightarrow  \e^{\I \gauge(x)\tau_z/2} \, \left| \Xi_j \right>$ and, thus, can be used in the 
original gauge.

\section{Spectrum}\label{appSpectrum}

The eigenmodes of the single particle Hamiltonian $h$, defined in Eqs.~\eqref{NambuH}, can be assembled from the piecewise solutions in the three different regions. 
Due to translation invariance, the $y$ dependence of all modes is given by plain waves $\exp(\I \, q \, y / \hbar)$ with momentum~$q$. 
In contrast, the solutions generally involve different momenta~$k$ along the $x$ axis, including imaginary values of $k$ for states that are localized in $x$ direction around the junction. 
As negative-energy modes are just the particle--hole inverted counterparts of positive-energy solutions, $\Xi_{-j} = \sigma_y \tau_y \, \Xi_j^*$, it is sufficient only to consider positive energies, $\epsilon > 0$, in the following. 
In the central region, denoted here by a subscript M, we can distinguish electron modes,
\begin{subequations}\label{xiM}
\begin{align}
\Xi^\text{(e)}_{k_\text{M} q}(x, y)
	&= \frac{1}{\sqrt{(\epsilon_\text{M} + M_0)^2 + v^2 \, |k_\text{M} + \I \, q|^2}} \,
	\begin{bmatrix}
	\epsilon_\text{M} + M_0 \\
	v \, (k_\text{M} + \I \, q) \\
	0 \\
	0
	\end{bmatrix} \, \e^{\I \, k_\text{M} x / \hbar} \, \e^{\I \, q \, y / \hbar}~,
\intertext{and hole modes of the same energy $\epsilon_\text{M} = \sqrt{M_0^2 + v^2 \, \big( k_\text{M}^2 + q^2 \big)}$~,}
\Xi^\text{(h)}_{k_\text{M} q}(x, y)
	&= \frac{1}{\sqrt{(\epsilon_\text{M} + M_0)^2 + v^2 \, |k_\text{M} + \I \, q|^2}} \,
	\begin{bmatrix}
	0 \\
	0 \\
	\epsilon_\text{M} + M_0 \\
	-v \, (k_\text{M} + \I \, q)
	\end{bmatrix} \, \e^{\I \, k_\text{M} x / \hbar} \, \e^{\I \, q \, y / \hbar}~.
\end{align}
\end{subequations}
Note that $k_\text{M}$ can be real or imaginary.
The superconducting coupling mixes electron and hole components, which yields the solutions
\begin{subequations}\label{xiS}
\begin{align}
\Xi^\text{($j$1)}_{k_j q}(x, y) &= \frac{1}{\sqrt{\epsilon_j^2 + \Delta_0^2 + v^2 \, |k_j + \I \, q|^2}} \,
	\begin{bmatrix}
	\Delta_0 \, \e^{-\I \, \phi_j} \\
	0 \\
	\epsilon_j \\
	-v \, (k_j + \I \, q)
	\end{bmatrix} \, \e^{\I \, k_j x / \hbar} \, \e^{\I \, q \, y / \hbar}~, \\
\Xi^\text{($j$2)}_{k_j q}(x, y) &= \frac{1}{\sqrt{\epsilon_j^2 + \Delta_0^2 + v^2 \, |k_j - \I \, q|^2}} \,
	\begin{bmatrix}
	v \, (k_j - \I \, q) \\
	\epsilon_j \\
	0 \\
	\Delta_0 \, \e^{\I \, \phi_j}
	\end{bmatrix} \, \e^{\I \, k_j x / \hbar} \, \e^{\I \, q \, y / \hbar}
\end{align}
\end{subequations}
for $\epsilon_j = \sqrt{\Delta_0^2 + v^2 \, \big( k_j^2 + q^2 \big)}$ in the left ($j =$ L) and right ($j =$ R) region. 
Note that $k_j$ can be real or imaginary.

Given fixed values of $q$ and $\epsilon$, only transverse momenta~$k_\text{M}, k_j$ producing the correct energy are relevant, \ie\ the solutions of the equations
\begin{align}
k_\text{M}^2 &= \frac{\epsilon^2 - M_0^2}{v^2} - q^2~, \\
k_j^2 &= \frac{\epsilon^2 - \Delta_0^2}{v^2} - q^2 = \begin{cases}
	k_\text{M}^2 - \frac{\Delta_0^2}{v^2}~, &\qquad \text{if } M_0 = 0~, \\
	k_\text{M}^2~, &\qquad \text{if } M_0 = \Delta_0~. \end{cases}
\end{align}
The Hamiltonian is linear in the momentum~$\pvec$, therefore the complete eigenmodes need to be continuous at the two boundaries $x = \pm W/2$.

\section{SNS case: Andreev bands}\label{SNSb}

Using the general approach of App.~\ref{appSpectrum}, we consider the Andreev bands of states localized in $x$ direction, first.  
In the SNS junction, Andreev states consist of propagating waves in the normal-conducting middle area, but decay into both of the superconducting regions. 
The transversal momentum in the middle region
\begin{align}
k_\text{M}^2 &\equiv  \frac{m^2}{v^2} \phantom{~.}
\end{align}
yields an effective mass term for the longitudinal wave in the energy:
\begin{align}
\epsilon &= \sqrt{m^2 + v^2 \, q^2}~. \label{eSNSb}
\end{align}
Although the transverse contribution to the energy has an upper bound $m^2 < \Delta_0^2$, the strict translation invariance in $y$~direction means that the Andreev bound states in this model exist alongside the continuum of scattering modes at energies $\epsilon \ge \Delta_0$, too. 
In the gapped regions, the momenta~$k_j$ are imaginary with the sign chosen to guarantee an exponential decay from the boundary,
\begin{align}
\I \, k_\text{L} &= +\frac{\sqrt{\Delta_0^2 - m^2}}{v}~, \\
\I \, k_\text{R} &= -\frac{\sqrt{\Delta_0^2 - m^2}}{v}~,
\end{align}
therefore the Andreev states have the form
\begin{align}
\Xi(x, y) &= \begin{cases}
	A_\text{(L1)} \, \Xi^\text{(L1)}_{k_\text{L} q}(x, y) + A_\text{(L2)} \, \Xi^\text{(L2)}_{k_\text{L} q}(x, y)~,
		& x < -W/2 \\[0.25cm]
	\sum_\pm \left[ A_\text{(e)}^\pm \, \Xi^\text{(e)}_{\pm k_\text{M} q}(x, y)
	+ A_\text{(h)}^\pm \, \Xi^\text{(h)}_{\pm k_\text{M} q}(x, y) \right]~,
		& |x| < W/2 \\[0.3cm]
	A_\text{(R1)} \, \Xi^\text{(R1)}_{k_\text{R} q}(x, y) + A_\text{(R2)} \, \Xi^\text{(R2)}_{k_\text{R} q}(x, y)~,
		& W/2 < x
	\end{cases}~.
\end{align}
The sign convention here is such that $k_\text{M} >0$. The inner structure of the Andreev bands is determined by the phenomenon of Andreev reflection at the SN boundaries. At the Dirac point, this reflection is specular, as first predicted in \Rcite{BeenakkerSpecular}.
The possible values for the mass~$m$ are restricted by the boundary conditions, as the wave function needs to be continuous at both boundary lines $x = \pm W/2$ for a (non-trivial) set of coefficients~$A$. 
This yields the quantization condition~\eqref{qSNSb}.

As we can calculate the current contribution of the Andreev bands from the $\phi$ dependence of the energies~\eqref{eSNSb} via the relation \eqref{qSNSb}, we do not need normalized wave functions here. 
For completeness, let us remark that the infinite-size limit in $y$ direction can be treated exactly like for the scattering states (see~App.~\ref{SNSsc}). 
Due to the condition~\eqref{qSNSb}, we retain a sum over discrete values of the transversal momentum, and the Andreev states are exponentially localized along $x$ anyway.

\section{SNS case: scattering states}\label{SNSsc}

Scattering states of fixed energy are conveniently characterized, \eg, by the longitudinal momentum~$q$ and the absolute value of the transversal momentum in the normal-conducting region, $|k_\text{M}| > \Delta_0 / v$:
\begin{align}
\Rightarrow \epsilon &= v \, \sqrt{k_\text{M}^2 + q^2}
\intertext{In all three regions (cf.~App.~\ref{appSpectrum}), waves can propagate in $\pm x$ direction, therefore the relevant transversal momenta are given by $\pm k_\text{M}$ in the middle region and}
k^2 &= k_\text{M}^2 - \Delta_0^2 / v^2
\end{align}
in the superconducting areas. 
Hence, a piece-wise representation of the the wave function is given by
\begin{align}
\Xi(x, y) &= \sum_\pm \begin{cases}
	A_\text{(L1)}^\pm \, \Xi^\text{(L1)}_{\pm k q}(x, y) + A_\text{(L2)}^\pm \, \Xi^\text{(L2)}_{\pm k q}(x, y)~,
		& x < -W/2 \\[0.25cm]
	A_\text{(e)}^\pm \, \Xi^\text{(e)}_{\pm k_\text{M} q}(x, y)
		+ A_\text{(h)}^\pm \, \Xi^\text{(h)}_{\pm k_\text{M} q}(x, y)~,
		& |x| < W/2 \\[0.25cm]
	A_\text{(R1)}^\pm \, \Xi^\text{(R1)}_{\pm k q}(x, y) + A_\text{(R2)}^\pm \, \Xi^\text{(R2)}_{\pm k q}(x, y)~,
		& W/2 < x
	\end{cases} \label{xiSNSsc}
\end{align}
for a suitable choice of coefficients~$A$. The sign convention here is such that $k_\text{M} >0$ and $k>0$.
Given fixed values of $k$ (or $k_\text{M}$) and $q$, there are four independent solutions fulfilling the condition of continuity at the boundaries $x = \pm W/2$. 
They correspond, \eg, to the four incoming waves $\Xi^\text{(L1)}_{+k q}, \Xi^\text{(L2)}_{+k q}, \Xi^\text{(R1)}_{-k q}$ and $\Xi^\text{(R2)}_{-k q}$. 
The boundary equations yield the transmission and reflection coefficients as well as the solutions in the ungapped middle. 

To ensure a correct normalization, we have to consider the system to be of finite dimensions $L_x$, $L_y$ at first. 
For the case of a single ingoing wave, the corresponding coefficient needs to be chosen as
\begin{align}
A_\text{in} &= \frac{1}{\sqrt{L_x \, L_y}}~.
\end{align}
In order to take the infinite-size limit, we have to replace sums over the momenta $k$, $q$ with integrals. 
For integrands quadratic in the coefficients~$A$, the lengths $L_x$ and $L_y$ cancel out in the end because they also appear in the integration measures. 
For scattering states (and Andreev states in the SNS case), the expression \eqref{eigen.Ixi} for the partial current  
of a single scattering state is identical to the expression
\begin{align}
I_j &= \frac{e \, v^2 \, k_\text{M}}{\epsilon} \int\! \td y
	\, \Bigl( |A_\text{(e)}^+|^2 - |A_\text{(e)}^-|^2 - |A_\text{(h)}^+|^2 + |A_\text{(h)}^-|^2 \Bigr) \label{eigen.IA}
\end{align}
in terms of the momentum~$k_\text{M}$, the energy~$\epsilon$ and the coefficients~$A$ of the eigenfunctions~\eqref{xiM} corresponding to the single-particle state $j$. The scattering-state contribution to the Josephson current for momenta~$k$, $q$ is calculated by summing the expression~\eqref{eigen.IA} for the four incoming-wave solutions:
\begin{align}
I_\text{sc}(k, q) &= e \, v \, \sqrt{\frac{\Delta_0^2 + v^2 \, k^2}{\Delta_0^2 + v^2 \, (k^2 + q^2)}}
	\, L_y \, \sum_{m = 1}^4 \Bigl( |A_\text{(e)}^+|^2 - |A_\text{(e)}^-|^2
	- |A_\text{(h)}^+|^2 + |A_\text{(h)}^-|^2 \Bigr)_{m, k q}~,
\intertext{where an additional factor of the system size~$L_y$ comes from the translation-invariant $y$ integral. 
This yields}
\frac{I_\text{sc}(k, q)}{L_y} &= \frac{2 \, e \, v}{L_x \, L_y}
	\, \sqrt{\frac{\Delta_0^2 + v^2 \, k^2}{\Delta_0^2 + v^2 \, (k^2 + q^2)}} \nn
&\quad\: \cdot \frac{\Delta_0^2 \; v^2 \, k^2 \, \sin\lft( 2 \, \frac{W}{\hbar}
	\, \sqrt{k^2 + \frac{\Delta_0^2}{v^2}} \right) \, \sin(\phi)}
	{\left[ v^2 \, k^2 + \Delta_0^2 \, \sin^2\lft( \frac{W}{\hbar} \, \sqrt{k^2 + \frac{\Delta_0^2}{v^2}}
	- \dfrac{\phi}{2} \right) \right] \left[ v^2 \, k^2 + \Delta_0^2 \, \sin^2\lft(
	\frac{W}{\hbar} \, \sqrt{k^2 + \frac{\Delta_0^2}{v^2}} + \dfrac{\phi}{2} \right) \right]}~.
\label{IkqSNS}
\end{align}

\section{SMS case: Andreev bands}\label{SMSb}

Due to the magnetic gap in the middle region, the transversal momentum~$k$ for the Andreev states is purely imaginary in all three parts of the system (cf.~App.~\ref{appSpectrum}), unlike in the SNS case of App.~\ref{SNSb}. 
We use the effective mass~$m < \Delta_0$ in the energy
\begin{align}
\epsilon &= \sqrt{m^2 + v^2 \, q^2} \label{eSMSb}
\intertext{to define the transverse momentum  $k$ for the solution decaying to the left:} 
 k &\equiv - \I \,\sqrt{\Delta_0^2 - m^2}~.
\end{align}
Andreev-state wave functions are given by
\begin{align}
\Xi(x, y) &= \begin{cases}
	A_\text{(L1)} \, \Xi^\text{(L1)}_{+k q}(x, y) + A_\text{(L2)} \, \Xi^\text{(L2)}_{+k q}(x, y)~,
		& x < -W/2 \\[0.25cm]
	\sum_\pm \left[ A_\text{(e)}^\pm \, \Xi^\text{(e)}_{\pm k q}(x, y)
	+ A_\text{(h)}^\pm \, \Xi^\text{(h)}_{\pm k q}(x, y) \right]~,
		& |x| < W/2 \\[0.3cm]
	A_\text{(R1)} \, \Xi^\text{(R1)}_{-k q}(x, y) + A_\text{(R2)} \, \Xi^\text{(R2)}_{-k q}(x, y)~,
		& W/2 < x
	\end{cases}~.
\end{align}

Like in the SNS case of Sec.~\ref{SNSb}, there are Andreev bound states in the continuum for large values of the longitudinal momentum~$q$, and the continuity constraint on the wave function yields a condition for the mass~$m$
given by Eq.~\eqref{qSMSb}.

\section{SMS case: Scattering states}\label{SMSsc}

The treatment of the scattering states is similar to the SNS case (App.~\ref{SNSsc}). 
In the magnetic region, the transversal momenta are the same as in the superconducting areas, so we can simply replace $k_\text{M} \rightarrow k$ in the definition of the wave function~\eqref{xiSNSsc}. 
Additionally, we need to take into account that the energy is given by
\begin{align}
\epsilon &= \sqrt{\Delta_0^2 + v^2 \, (k^2 + q^2)}
\end{align}
for a state with longitudinal momentum~$q$ and absolute value of the transversal momentum~$k$. 

By summing the expression~\eqref{eigen.IA} for the four incoming-wave solutions,
\begin{align}
I_\text{sc}(k, q) &= \frac{e \, v^2 \, k}{\sqrt{\Delta_0^2 + v^2 \, (k^2 + q^2)}}
	\, L_y \, \sum_{m = 1}^4 \Bigl( |A_\text{(e)}^+|^2 - |A_\text{(e)}^-|^2
	- |A_\text{(h)}^+|^2 + |A_\text{(h)}^-|^2 \Bigr)_{m, k q}
\intertext{we obtain the expression}
\frac{I_\text{sc}(k, q)}{L_y} &= \frac{2 \, e \, v}{L_x \, L_y}
	\, \frac{v \, k}{\sqrt{\Delta_0^2 + v^2 \, (k^2 + q^2)}} \,
	\frac{\Delta_0^2 \; \big( \Delta_0^2 + v^2 \, k^2 \big) \, \sin[2 \, W \, k / \hbar] \, \sin[\phi]}
	{\Bigl( v^2 \, k^2 + \Delta_0^2 \, \sin^2\lft[ \frac{\phi}{2} \right] \Bigr)^2 + \, 4 \, \Delta_0^2 \;
	\big( \Delta_0^2 + v^2 \, k^2 \big) \, \cos^2\lft[ \frac{\phi}{2} \right] \, \sin^2[W \, k / \hbar]}~.
	\label{IkqSMS}
\end{align}

\bibliography{jj2d}

\end{document}